\documentclass[3p]{elsarticle}
\usepackage{amssymb}
\usepackage{amsmath}
\usepackage{graphicx}
\usepackage{color}
\usepackage{float}

\begin{document}

\begin{frontmatter}
\title{On the quasistatic effective elastic moduli for
elastic waves in three-dimensional phononic crystals}

\author[aak]{A.A. Kutsenko}
\ead{aak@nxt.ru}
\author[aak]{A.L. Shuvalov}
\ead{a.shuvalov@i2m.u-bordeaux1.fr}
\author[ann]{A.N. Norris\corref{cor1}}
\ead{norris@rutgers.edu}

\cortext[cor1]{Corresponding author}

\address[aak]{Institut de M\'{e}canique et d'Ing\'{e}nierie de Bordeaux, \\
Universit\'{e} de Bordeaux, UMR CNRS 5469, Talence 33405, France}
\address[ann]{Mechanical and Aerospace Engineering, Rutgers University,\\
Piscataway, NJ 08854-8058, USA}

\begin{abstract}

Effective elastic moduli for 3D solid-solid phononic crystals of arbitrary anisotropy and  oblique lattice structure are formulated analytically using the plane-wave expansion (PWE) method and  the recently proposed  monodromy-matrix (MM) method.  The latter approach employs Fourier series in two dimensions with direct numerical integration along the third direction.  As a result, the MM method converges much  quicker to the exact moduli in comparison with the PWE as the number of Fourier coefficients increases.    The MM method yields a more explicit formula than previous results, enabling a  closed-form upper bound on the effective Christoffel tensor. The MM approach significantly improves the efficiency and accuracy of evaluating effective wave speeds for high-contrast composites and for configurations of closely spaced inclusions, as demonstrated by three-dimensional examples.
\end{abstract}
\end{frontmatter}

\section{Introduction} \label{sec1}

Long-wave low-frequency dispersion of acoustic waves in periodic structures
is of both fundamental and practical interest, particularly due to the
current advances in manufacturing of metamaterials and phononic crystals. In
this light, the leading order dispersion (quasistatic limit) has recently
been under intensive study by various theoretical approaches such as
plane-wave expansion (PWE) \cite{Krokhin03,Ni05,KSNP}, scaling technique \cite%
{Parnell08b,AADW}, asymptotics of multiple-scattering theory \cite{Sheng07,Torrent08,Wu09} and a
newly proposed monodromy-matrix (MM) method \cite{Kutsenko11}.
The cases treated were mostly confined to scalar waves in 2D (two-dimensional)
structures.
{Regarding vector waves in 2D and especially in 3D phononic
crystals,
a variety of methods have been proposed for calculating the quasistatic effective elastic properties of  3D periodic composites containing spherical inclusions  arranged in  a simple cubic array.  For the case of
rigid inclusions an integral equation on  the sphere surface was solved numerically to obtain the effective properties \cite{Nunan84}.  Spherical voids \cite{Nasser81} and subsequently elastic inclusions
were considered using a Fourier series approach \cite{Nasser82}. Alternative procedures for elastic spherical inclusions include the method of singular distributions \cite{Sangani87} and  infinite series of periodic multipole solutions of the equilibrium equations \cite{Kusch87}.
The latter multipole expansion method has also been applied to cubic arrays of ellipsoidal inclusions  \cite{Kusch97}.  A particular
PWE-based method of calculation of quasistatic speeds in 3D phononic
crystals of cubic symmetry has been formulated and implemented in \cite{Ni07}.
A review of numerical methods for calculating effective properties of composites  can be found in \cite[\S2.8,\S14.11]{Milton01}.
}


{Of all the methods available for calculating effective elastic
moduli the PWE method is arguably the simplest and most
straightforward to implement.  It requires only Fourier coefficients
of the inclusion in the unit cell, which makes it the method of
choice for many problems.} Unfortunately, PWE is not a very
practical tool for the 3D case, where the vectors and matrices in
the Fourier space are of very large algebraic dimension, especially
if the phononic crystal is composed of highly contrasting materials
(examples  in \S\ref{sec5} illustrate this critical  drawback).

The present paper provides the PWE and MM analytical formulations of the 21 components of the  effective elastic stiffness
for 3D solid-solid phononic crystals of arbitrary
anisotropy and arbitrary oblique lattice.   While the PWE method is widely used in some fields its formulation for general anisotropic static elasticity has, surprisingly, not been discussed  before.  The PWE is presented here in a compact form (see Eq.\  \eqref{PWEF}) suitable for numerical implementation.
The main thrust of the paper is concerned with the MM approach. The motivation for advocating this method as an alternative to the more conventional
PWE technique is, first, that the MM method 'spares' Fourier expansion in
one of the coordinates (this is particularly advantageous for the 3D
numerics) and, second, that the MM method has much faster convergence than
PWE. Comparison of the MM and PWE calculations provided in the paper
confirms a markedly better efficiency of the MM method.

The paper is organised as follows. In Section \ref{sec2}, the quasistatic
perturbation theory is used to define the effective Christoffel equation in
the form which serves as the  common starting point for the PWE and MM methods.
The PWE formulation of the effective elastic moduli follows readily and is also presented in Section \ref{sec2}. The MM formulation is described in Section \ref{sec3}: the
derivation of the MM formula is in \S\ref{sec3.1} (see also Appendix 1),
 its numerical implementation is discussed in \S\ref{sec3.2},
 generalization to the case of an oblique lattice is presented in \S\ref{3.4}
  and the scheme for recovering the full set of effective elastic moduli is provided in \S\ref{full}.  A closed-form
estimate of the effective Christoffel matrix is presented in Section \ref{sec4}.
Examples of the MM and PWE calculations are provided in Section \ref{sec5}.
Concluding remarks are given in Section \ref{sec6}.

\section{Background. PWE formula}\label{sec2}

Consider a 3D anisotropic medium with density and elastic stiffness
\begin{equation}
\rho (\mathbf{x})=\rho (\mathbf{x+e}_{p}),\ 
{\mathbf{C}(\mathbf{x}) =\mathbf{C}(\mathbf{x+e}_{p}),}
  \label{-1}
\end{equation}%
which are assumed to be $\mathbf{1}$-periodic, i.e. invariant to period or
translation vectors $\mathbf{e}_{p}=\left( \delta _{pq}\right) $ (a cubic
lattice, otherwise see \S\ref{3.4}). All roman indices run from 1 to 3. In
the following,
$^{\ast }$ and $^{+}$ mean complex and Hermitian conjugation.
Assume no dissipation so that {the elements of $\mathbf{C}$ satisfy} $c_{ijkl}=c_{klij}^{\ast }~(=c_{klij}$ for
real case).  Our   goal is the  quasistatic effective elastic stiffness $\mathbf{C}^{\mathrm{eff}}$
with elements $c_{ijkl}^{\mathrm{eff}}$ that have the same symmetries as  those of
$\mathbf{C}$, and matrices  $\mathbf{C}_{jl}^{\mathrm{eff}}$ defined by analogy with Eq.\ \eqref{0}.
For  compact writing, introduce the matrices
\begin{equation}
\mathbf{C}_{jl}=\left( c_{ijkl}\right) _{i,k=1}^{3}=\mathbf{C}_{lj}^{+}
\label{0}
\end{equation}%
with components numbered by $i,k$.
The elastodynamic
equation for time-harmonic waves $\mathbf{v}\left( \mathbf{x},t\right) =%
\mathbf{v}( \mathbf{x} ) e^{-i\omega t}$ is%
\begin{equation}
\partial _{j}(\mathbf{C}_{jl}\partial _{l}\mathbf{v})=-\rho \omega ^{2}%
\mathbf{v},\   \label{1}
\end{equation}%
where $\partial _{j}\equiv \partial /\partial x_{j}$ and repeated indices
are summed.
The differential operator in Eq.\ (\ref{1}) is self-adjoint with
respect to the Floquet condition $\mathbf{v}(\mathbf{x})=\mathbf{u}(\mathbf{x%
})e^{i\mathbf{k\cdot x}}$ with $\mathbf{1}$-periodic $\mathbf{u}(\mathbf{x})=%
\mathbf{u}(\mathbf{x+e}_{p})$ and $\mathbf{k}=k\pmb{\kappa }\ \ (\left\vert
\pmb{\kappa
}\right\vert =1)$. Substituting this condition in (\ref{1}) casts it into the
form%
\begin{equation}
\begin{aligned}
& (\mathcal{C}_{0}+k\mathcal{C}_{1}+k^{2}\mathcal{C}_{2})\mathbf{u}=\rho
\omega ^{2}\mathbf{u},\ \ \mathrm{where\ }
\\
\mathcal{C}_{0}\mathbf{u}\equiv -\partial _{j}(\mathbf{C}_{jl}\partial _{l}%
\mathbf{u}),\  & \mathcal{C}_{1}\mathbf{u}\equiv -i(\kappa _{j}\mathbf{C}%
_{jl}\partial _{l}+\kappa _{l}\partial _{j}\mathbf{C}_{jl})\mathbf{u},\
\mathcal{C}_{2}\mathbf{u}\equiv \kappa _{j}\kappa _{l}\mathbf{C}_{jl}\mathbf{%
u}.%
\end{aligned}
\label{we1}
\end{equation}
All operators $\mathcal{C}$ are self-adjoint. {We introduce for
future use linear, areal and volumetric averages over the unit-cell:
$\langle \cdot\rangle _{j}$ is  the average over coordinate $x_j$;
$\langle \cdot\rangle_{\overline{j}}$ is the average over the
section orthogonal to $x_{j}$,  and $\langle \cdot\rangle$ is the
complete average.  These averages in turn define inner products of
vector-valued functions. } Thus, for  a scalar function $f$ and
vector-functions $\mathbf{f,}$ $\mathbf{h}$,
\begin{equation}\label{000}
\langle f\rangle _{j}=\int_{0}^{1}fdx_{j},\ \
\langle f\rangle
=
\int_{[0,1]^{3}}fd\mathbf{x},
\ \  {
(\mathbf{f},\mathbf{h})_{\overline{j}}=
\langle \mathbf{h}^{+}\mathbf{f}\rangle_{\overline{j}},} \ \
(\mathbf{f},\mathbf{h})=\langle \mathbf{h}^{+}\mathbf{f}\rangle .
\end{equation}%

Next we apply  perturbation theory to (\ref{we1}) with a view to defining
the quasistatic effective Christoffel matrix whose eigenvalues yield the
effective speeds\textit{\ }
\begin{equation}
c_{\alpha }\equiv c_{\alpha }(\pmb{\kappa})\equiv \lim_{k\rightarrow
0}\omega _{\alpha }\left( \mathbf{k}\right) /k,\ \alpha =1,2,3.  \label{c}
\end{equation}%
For $k=0,$ the eigenvalue $\omega =0$ has multiplicity $3$ and corresponds to
three constant linear independent eigenvectors $\mathbf{u}_{0\alpha }$.
Consider the asymptotics
\begin{subequations}\label{as}
\begin{align}
\omega _{\alpha }^{2} &=0+k\lambda _{1\alpha }+k^{2}\lambda _{2\alpha
}+O(k^{3}),   \\
\mathbf{u}_{\alpha } &=\mathbf{u}_{0\alpha }+k\mathbf{u}_{1\alpha }+k^{2}%
\mathbf{u}_{2\alpha }+\mathbf{O}(k^{3}).
\end{align}%
\end{subequations}
Substituting (\ref{as}) into (\ref{we1}) and collecting terms with the same
power of $k$ yields
\begin{subequations}
\begin{align}
1 &:  & \mathcal{C}_{0}\mathbf{u}_{0\alpha }
&=\mathbf{0,} \\
k &: & \mathcal{C}_{1}\mathbf{u}_{0\alpha }+\mathcal{C}_{0}\mathbf{u}_{1\alpha
}
&=\rho \lambda _{1\alpha }\mathbf{u}_{0\alpha },  \label{ass1} \\
k^{2} &: & \mathcal{C}_{2}\mathbf{u}_{0\alpha }+\mathcal{C}_{1}\mathbf{u}%
_{1\alpha }+\mathcal{C}_{0}\mathbf{u}_{2\alpha }
&=\rho \lambda _{1\alpha }%
\mathbf{u}_{1\alpha }+\rho \lambda _{2\alpha }\mathbf{u}_{0\alpha }.
\qquad\qquad\qquad
\label{ass2}
\end{align}%
\end{subequations}
Scalar multiplying (\ref{ass1}) by $\mathbf{u}_{0\alpha }$ and (\ref{ass2})
by $\mathbf{e}_{k},$ and using $(\mathcal{C}_{1}\mathbf{u}_{0\alpha },%
\mathbf{u}_{0\alpha })=0$ together with self-adjointness of $\mathcal{C}_0$
leads to
\begin{align}  \label{ass3}
\lambda _{1\alpha } =0;\ \ \mathbf{u}_{1\alpha }=-\mathcal{C}_{0}^{-1}%
\mathcal{C}_{1}\mathbf{u}_{0\alpha };   \ \
(\mathcal{C}_{2}\mathbf{u}_{0\alpha },\mathbf{e}_{k})+(\mathcal{C}_{1}%
\mathbf{u}_{1\alpha },\mathbf{e}_{k}) =\langle \rho \rangle \lambda
_{2\alpha }(\mathbf{u}_{0\alpha },\mathbf{e}_{k}),\ k=1,2,3,  
\end{align}%
where $\lambda _{2\alpha }=c_{\alpha }^{2}$ due to $\lambda _{1\alpha }=0$
and (\ref{c}).\textit{\ }Inserting $\mathbf{u}_{1\alpha }$ from (\ref{ass3})$_2$
in (\ref{ass3})$_3$ gives
\begin{equation}
{\boldsymbol{\Gamma}}\mathbf{\mathbf{u}}_{0\alpha }=\langle \rho \rangle c_{\alpha
}^{2}\mathbf{\mathbf{u}}_{0\alpha },\ \ \alpha =1,2,3,  \label{eq}
\end{equation}%
where ${\boldsymbol{\Gamma}}=((\mathcal{C}_{2}\mathbf{e}_{k},\mathbf{e}_{i})-(%
\mathcal{C}_{0}^{-1}\mathcal{C}_{1}\mathbf{e}_{k},\mathcal{C}_{1}\mathbf{e}_{i}))_{i,k=1}^{3}.$ Substituting $\mathcal{C}_{1},~\mathcal{C}_{2}$ from (%
\ref{we1}) defines the quasistatic effective $3\times 3$ Christoffel matrix $%
{\boldsymbol{\Gamma}}=\left( \Gamma _{ik}\right) _{i,k=1}^{3}$ in the form
\begin{equation} \label{em}
{\boldsymbol{\Gamma}}( \mathbf{\kappa })
=\mathbf{C}_{jl}^{\mathrm{e}}\kappa _{j}\kappa _{l}
\ \ \text{where}\ \
 \mathbf{C}_{jl}^{\mathrm{e}}
\equiv \langle \mathbf{%
C}_{jl}\rangle -\langle (\partial _{p}\mathbf{C}_{pj}^{+})\mathcal{C}%
_{0}^{-1}(\partial _{q}\mathbf{C}_{ql})\rangle
\  \big( =  \mathbf{C}_{lj}^{\mathrm{e}^+} \big).
\end{equation}
{The matrix $ \mathbf{C}_{jl}^{\mathrm{e}}$ is distinguished from $
\mathbf{C}_{jl}^{\mathrm{eff}}$, in terms of which the Christoffel
matrix is
\begin{equation} \label{9-0}
{\boldsymbol{\Gamma}}( \mathbf{\kappa })
=\mathbf{C}_{jl}^{\mathrm{eff}}\kappa _{j}\kappa _{l}
=\frac 12 \big( \mathbf{C}_{jl}^{\mathrm{eff}}
+ \mathbf{C}_{lj}^{\mathrm{eff}} \big) \kappa _{j}\kappa _{l}
.
\end{equation}
Comparison of Eqs.\ \eqref{em}$_1$ and \eqref{9-0} implies that
$\mathbf{C}_{jl}^{\mathrm{e}}$ and $\mathbf{C}_{jl}^{\mathrm{eff}}$
are related by
\begin{equation} \label{123}
\mathbf{C}_{jl}^{\mathrm{eff}} +\mathbf{C}_{lj}^{\mathrm{eff}}   =
\mathbf{C}_{jl}^{\mathrm{e}} +\mathbf{C}_{lj}^{\mathrm{e}}
\end{equation}
and  they are equal if $j=l$.  Equation \eqref{em}$_2$  does not yield   $\mathbf{C}_{jl}^{\mathrm{eff}}$ explicitly,  only in the combination \eqref{123};   however, this connection is sufficient for the purpose of finding  all elements of  $\mathbf{C}^{\mathrm{eff}}$, as described in \S\ref{full}.  For now we focus on methods to  calculate $\mathbf{C}_{jl}^{\mathrm{e}}$ .
}

Equation  \eqref{em}$_2$ is still an implicit formula for the matrix  $\mathbf{C}_{jl}^{\mathrm{e}}$ in so far as the operator $%
\mathcal{C}_{0}^{-1}$ is not specified. One way to an explicit implementation
of \eqref{em}$_2$ is via its transformation to  Fourier space which must be
truncated to define $\mathcal{C}_{0}^{-1}$ as a  matrix inverse. Thus
taking the 3D Fourier expansion
\begin{equation}
\mathbf{C}_{jl}(\mathbf{x})=\sum\nolimits_{\mathbf{g}\in \mathbb{Z}^{3}}%
\widehat{\mathbf{C}}_{jl}(\mathbf{g})e^{2\pi i\mathbf{g}\cdot \mathbf{x}},\ \
\widehat{\mathbf{C}}_{jl}(\mathbf{g})=\langle \mathbf{C}_{jl}(\mathbf{x%
})e^{-2\pi i\mathbf{g}\cdot \mathbf{x}}\rangle  \label{fur3}
\end{equation}
and plugging it into \eqref{em}$_2$ yields the PWE formula for {$\mathbf{C}_{jl}^{\mathrm{e}}$} as follows
\begin{equation}\label{PWEF}
{\mathbf{C}_{jl}^{\mathrm{e}}}=\widehat{\mathbf{C}}_{jl}(\mathbf{0})-\mathbf{%
q}_{j}^{+}\mathbf{C}_{0}^{-1}\mathbf{q}_{l},\ \ \mathrm{where}
\ \  
\mathbf{q}_{j}=(g_{i}\widehat{\mathbf{C}}_{ij}(\mathbf{g}))_{\mathbf{g}\in
\mathbb{Z}^{3}\backslash \mathbf{0}},\ \ \mathbf{C}_{0}=(g_{k}g_{p}^{\prime }%
\widehat{\mathbf{C}}_{kp}(\mathbf{g}-\mathbf{g}^{\prime }))_{\mathbf{g},%
\mathbf{g}^{\prime }\in \mathbb{Z}^{3}\backslash \mathbf{0}}.
\end{equation}
This result corresponds to the quasistatic limit of the  effective
elastic coefficients for a  dynamically homogenized periodic medium, see \cite[eq.\ (2.11)]{Norris12}.

Unfortunately, although the PWE formula (\ref{PWEF}) is straightforward, its
numerical use in 3D is complicated by the large algebraic dimensions of vectors
and matrices of Fourier coefficients. This motivates using the monodromy
matrix (MM) method which confines the PWE computation to  2D, see next Section. The MM
method will also be shown to have  significantly faster convergence than PWE.

\section{MM formula}\label{sec3}

\subsection{Derivation}\label{sec3.1}

{The idea underlying the MM method is to reduce the 3-dimensional
problem of Eq.\  \eqref{em}$_2$ to an   equivalent 1-dimensional
equation that can be integrated. This is achieved by focusing on a
single  coordinate and using Fourier transforms in the orthogonal
coordinates.} The MM formula may be deduced proceeding from Eq.\
\eqref{em}$_2$; using integration by parts, let us recast it into
the form
\begin{equation}
{\mathbf{C}_{jl}^{\mathrm{e}}}=\langle \mathbf{C}_{jl}\rangle -\langle
(\partial _{p}\mathbf{C}_{pj}^{+})\mathbf{U}\rangle =
\langle \mathbf{C}_{jp}\partial _{p}\mathbf{V}%
\rangle ,  \label{eq1}
\end{equation}%
where we have denoted $\mathcal{C}_{0}^{-1}(\partial _{q}\mathbf{C}_{ql})=%
\mathbf{U}$ and $\mathbf{V=U}+x_{l}\mathbf{I}_{3}$ with $\mathbf{I}_{3}$
standing for the $3\times 3$ identity matrix. Thus we need to solve the equation
\begin{equation}\label{hh}
\mathcal{C}_{0}\mathbf{U}
=\partial _{q}\mathbf{C}_{ql}\ \ \Leftrightarrow \ \
\mathcal{C}_{0}\mathbf{U}=-\mathcal{C}_{0}x_{l}\mathbf{I}_{3}\ \
\Leftrightarrow
\ \
\mathcal{C}_{0}\mathbf{V}
=0\ \ \Leftrightarrow \ \ \partial _{q}(\mathbf{C}%
_{qp}\partial _{p}\mathbf{V})=0\ \mathrm{with}\ \mathbf{V}=\mathbf{U}+x_{l}%
\mathbf{I}_{3}.
\end{equation}%
In the following derivation the indices $j,l$ are regarded as fixed, all
repeated indices are summed and among them the indices $a,b$ are specialized
by the condition $a,b\neq l.$ The suffix $0$ of the differential operators  $%
\boldsymbol{\mathcal{Q}}_{0},~\boldsymbol{\mathcal{M}}_{0}$ below indicates their reference to $\omega
,k=0$ (similarly to $\mathcal{C}_{0}$).

Equation (\ref{hh}) can be rewritten {as an ordinary differential
equation in the designated coordinate $x_l$}
\begin{equation}\label{hh0}
\boldsymbol{\Xi}^{\prime }=\boldsymbol{\mathcal{Q}}_{0}\ \mathbf{\Xi }\ \ \mathrm{with}\ \
^{\prime }\equiv \partial _{l},    \ \
\mathbf{\Xi }=%
\begin{pmatrix}
\mathbf{V} \\
\mathbf{C}_{lp}\partial _{p}\mathbf{V}%
\end{pmatrix}%
,\ \boldsymbol{\mathcal{Q}}_{0}=%
\begin{pmatrix}
-\mathcal{B}^{-1}\mathcal{A}_{1} & \mathcal{B}^{-1} \\
\mathcal{A}_{2}-\mathcal{A}_{1}^{+}\mathcal{B}^{-1}\mathcal{A}_{1} &
\mathcal{A}_{1}^{+}\mathcal{B}^{-1}%
\end{pmatrix}%
,
\end{equation}%
where $\mathbf{\Xi }$ is a $6\times 3$ matrix and the matrix operators $%
\mathcal{B}$, $\mathcal{A}_{1}$ and $\mathcal{A}_{2}$ are defined by
\begin{equation}
\mathcal{B}\mathbf{V}=\mathbf{C}_{ll}\mathbf{V},\ \ \mathcal{A}_{1}\mathbf{V}%
=\mathbf{C}_{lb}\partial _{b}\mathbf{V},\ \mathcal{A}_{2}\mathbf{V}%
=-\partial _{a}(\mathbf{C}_{ab}\partial _{b}\mathbf{V}).  \label{hh2}
\end{equation}%
Note that $\mathcal{B}$ and $\mathcal{A}_{2}$ are self-adjoint at any fixed $x_{l}$ {with respect to the inner product $(\cdot , \cdot )_{\overline{l}}$ (see \eqref{000}).}
 Denote
$\mathbf{\Xi }( s) \equiv \mathbf{\Xi }( \mathbf{x} )
\mid_{x_{l}=s}$.
The solution to (\ref{hh0}) with some initial matrix function $\mathbf{\Xi }%
\left( 0\right) $ has the form
\begin{equation}
\mathbf{\Xi }(x_{l})=\boldsymbol{\mathcal{M}}_{0}(x_{l})
\, \mathbf{\Xi }\left( 0\right) \ \
\mathrm{with\ }\boldsymbol{\mathcal{M}}_{0}(x_{l})=\widehat{\int _{0}^{x_{l}} }(\boldsymbol{\mathcal{I}}%
+\boldsymbol{\mathcal{Q}}_{0}dx_{l}),  \label{monhh}
\end{equation}%
where ${\boldsymbol{\mathcal{M}}}_{0}$ is a propagator matrix  defined through the multiplicative
integral $\widehat{\int }$ and $\boldsymbol{\mathcal{I}}$ is the identity operator. Note
the identities%
\begin{equation}
\mathrm{for\ }\mathbf{W}_{0}=%
\begin{pmatrix}
\mathbf{I}_{3} \\
\mathbf{0}%
\end{pmatrix}%
,\ \widetilde{\mathbf{W}}_{0}=%
\begin{pmatrix}
\mathbf{0} \\
\mathbf{I}_{3}%
\end{pmatrix} : \ \
\boldsymbol{\mathcal{Q}}_{0}\mathbf{W}_{0}
=\mathbf{0,\ }\   \boldsymbol{\mathcal{Q}}_{0}^{+}\widetilde{%
\mathbf{W}}_{0}=\mathbf{0}\
\ \
\Rightarrow  \ \
\boldsymbol{\mathcal{M}}_{0}\mathbf{W}_{0}
=\mathbf{W}_{0},\   \boldsymbol{\mathcal{M}}_{0}^{+}\widetilde{%
\mathbf{W}}_{0}=\widetilde{\mathbf{W}}_{0}%
.
\label{i}
\end{equation}

It is seen that for any value of $x_{l}$ the operator $\boldsymbol{\mathcal{M}}_{0}-\boldsymbol{\mathcal{I}}$
has no inverse but at the same time it is a one-to-one mapping from the
subspace orthogonal to $\mathbf{W}_{0}$ onto the subspace orthogonal to $%
\widetilde{\mathbf{W}}_{0}.$ By the definitions of $\mathbf{V}$ and $\mathbf{%
\Xi }$ in (\ref{hh}), (\ref{hh0}) and due to periodicity of $\mathbf{U,}$ it
follows that
\begin{equation}
\mathbf{\Xi }(1)=\mathbf{\Xi }(0)+\mathbf{W}_{0}\ \Rightarrow \ \mathbf{\Xi }%
(0)=(\boldsymbol{\mathcal{M}}_{0}(1)-\boldsymbol{\mathcal{I}})^{-1}\mathbf{W}_{0}\equiv \mathbf{S}.
\label{W}
\end{equation}%
Thus from (\ref{monhh}), (\ref{W}) and (\ref{hh0}), (\ref{i}),
\begin{equation}
\mathbf{\Xi}(x_{l})=\boldsymbol{\mathcal{M}}_{0}(x_{l})\mathbf{S}\ \ \Rightarrow \ \mathbf{V=W}_{0}^{+}\boldsymbol{\mathcal{M}}_{0}(x_{l})\mathbf{S},\ \ \mathbf{C}_{lp}\partial _{p}\mathbf{V}=%
\widetilde{\mathbf{W}}_{0}^{+}\boldsymbol{\mathcal{M}}_{0}(x_{l})\mathbf{S}
. \label{eta1}
\end{equation}
Substituting  $\mathbf{V}$  obtained from \eqref{eta1} into
(\ref{eq1}) yields the desired formula for
{$\mathbf{C}_{jl}^{\mathrm{e}}$}, {which is discussed further in
\S\ref{3.2.3}}. {Note, since  $\boldsymbol{\mathcal{M}}$ is
self-adjoint with respect to   $(\cdot , \cdot )_{\overline{l}}$,
\begin{equation}\label{321}
\langle \mathbf{C}_{lp}\partial _{p}\mathbf{V} \rangle_{\overline{l}}
= ( \boldsymbol{\mathcal{M}}_{0}(x_{l})\mathbf{S},
\widetilde{\mathbf{W}}_{0} )_{\overline{l}}
= ( \mathbf{S}, \boldsymbol{\mathcal{M}}_{0}^+(x_{l})
\widetilde{\mathbf{W}}_{0} )_{\overline{l}}
= ( \mathbf{S}, \widetilde{\mathbf{W}}_{0} )_{\overline{l}}
= \langle  \widetilde{\mathbf{W}}_{0}^{+}\mathbf{S}\rangle_{\overline{l}} .
\end{equation}
The latter identity implies that the laterally averaged  'traction'
component of $\mathbf{\Xi }(x_{l})$ is independent of $x_{l}$, and may be identified as a   net 'force' acting on the faces $x_{l}$= constant. }

Further simplification can be achieved for the matrices $\mathbf{C}_{ll}^{\mathrm{eff}}$
{$(=\mathbf{C}_{ll}^{\mathrm{e}} )$}.
By (\ref{eq1}) and (\ref{321})
\begin{equation}
\mathbf{C}_{ll}^{\mathrm{eff}}=\langle \mathbf{C}_{lp}\partial _{p}\mathbf{V}%
\rangle =
{\langle \widetilde{\mathbf{W}}_{0}^{+}\mathbf{S}\rangle
}
\equiv \langle \widetilde{\mathbf{W}}_{0}^{+}(\boldsymbol{\mathcal{M}}_{0}(1)-\boldsymbol{\mathcal{I}})^{-1}\mathbf{W}_{0}\rangle _{\overline{l}}.  \label{hh3}
\end{equation}%
Equation (\ref{hh3}) suffices to
define the effective Christoffel tensor for $\mathbf{k}=k\mathbf{\kappa }$
parallel to the translation vector $\mathbf{e}_{l}$ in which case ${%
\boldsymbol{\Gamma}}( \mathbf{\kappa }) =\mathbf{C}_{ll}^{\mathrm{eff}}$%
. Note that an alternative derivation of (\ref{hh3}) is given in Appendix 1.

Finally it is noted that $\boldsymbol{\mathcal{M}}_{0}(1)$ which appears in the above
expressions is formally a monodromy matrix (MM) relatively to the coordinate
$x_{l}$, for which reason this approach and its outcome formulas are
referred to as the MM ones. The MM approach to finding effective speed of
shear (scalar) waves in 2D structures was first  presented in \cite{Kutsenko11}%
, where Eq.\ (\ref{hh3}) for vector waves was also mentioned but with neither
derivation nor discussion\textit{.}

\subsection{Implementation}\label{sec3.2}

\subsubsection{Propagator matrix in direction $\mathbf{e}_{l}$}

As above, we fix $l$ and keep $a,b\neq l$. Introduce the 2D Fourier expansion
\begin{equation}
\mathbf{C}_{pq}(\mathbf{x})=\sum\nolimits_{\mathbf{g}\in \mathbb{Z}%
^{2}}e^{2\pi ig_{a}x_{a}}\widehat{\mathbf{C}}_{pq}(\mathbf{g},x_{l}),\ \
\widehat{\mathbf{C}}_{pq}(\mathbf{g},x_{l})=\langle \mathbf{C}_{pq}(\mathbf{x%
})e^{-2\pi ig_{a}x_{a}}\rangle _{\overline{l}}.  \label{fur2}
\end{equation}%
Operators $\mathcal{B},\mathcal{A}_{1},\mathcal{A}_{2}$ and matrix operators
$\boldsymbol{\mathcal{Q}},\boldsymbol{\mathcal{M}}_{0}(x_{l})$ defined in (\ref{hh0})-(\ref{monhh})
are represented in the 2D Fourier space by the following infinite matrices
truncated to a finite size in calculations:
\begin{subequations}
\begin{align}
\qquad \qquad \qquad \qquad
\mathcal{B}\ & \mapsto \ &\mathbf{B} &=(\widehat{\mathbf{C}}_{ll}(\mathbf{g}-%
\mathbf{g}^{\prime },x_{l}))_{\mathbf{g},\mathbf{g}^{\prime }\in \mathbb{Z}%
^{2}},
\qquad \qquad \qquad \qquad \qquad
\label{matr1} \\
\mathcal{A}_{1}\ &\mapsto \ &\mathbf{A}_{1} &=2\pi i(\widehat{\mathbf{C}}%
_{lb}(\mathbf{g}-\mathbf{g}^{\prime },x_{l})g_{b}^{\prime })_{\mathbf{g},%
\mathbf{g}^{\prime }\in \mathbb{Z}^{2}},  \label{matrA1} \\
\mathcal{A}_{2}\ &\mapsto \ &\mathbf{A}_{2} &=4\pi ^{2}(g_{a}\widehat{\mathbf{%
C}}_{ab}(\mathbf{g}-\mathbf{g}^{\prime },x_{l})g_{b}^{\prime })_{\mathbf{g},%
\mathbf{g}^{\prime }\in \mathbb{Z}^{2}},  \label{matrA2} \\
\boldsymbol{\mathcal{Q}}_{0}\ &\mapsto \ &\mathbf{Q}_{0} &=%
\begin{pmatrix}
-\mathbf{B}^{-1}\mathbf{A}_{1} & \mathbf{B}^{-1} \\
\mathbf{A}_{2}-\mathbf{A}_{1}^{+}\mathbf{B}^{-1}\mathbf{A}_{1} & \mathbf{A}%
_{1}^{+}\mathbf{B}^{-1}%
\end{pmatrix}%
,  \label{matrQ} \\
\boldsymbol{\mathcal{M}}_{0}(x_{l})\ &\mapsto \ &\mathbf{M}_{0}(x_{l}) &=\widehat{\int_{0}^{x_{l}}}(\mathbf{I}+\mathbf{Q}_{0}dx_{l}).  \label{matr2}
\end{align}
\end{subequations}

\subsubsection{Principal directions $\mathbf{\protect\kappa }\parallel
\mathbf{e}_{l}$}
Consider the effective matrices $\mathbf{C}_{ll}^{\mathrm{eff}}$~($=\mathbf{%
\Gamma }$ if$\ \mathbf{\kappa }\parallel \mathbf{e}_{l}$). In view of the
above notations, formula (\ref{hh3}) for $\mathbf{C}_{ll}^{\mathrm{eff}}$ is
expressed as%
\begin{equation}
\mathbf{C}_{ll}^{\mathrm{eff}}=\widetilde{\mathbf{W}}_{\widehat{0}}^{+}%
\widehat{\mathbf{S}}\ \ \mathrm{with\ }\ \widehat{\mathbf{S}}=(\mathbf{M}%
_{0}(1)-\mathbf{I})^{-1}\mathbf{W}_{\widehat{0}},  \label{Cjjfur}
\end{equation}%
where%
\begin{equation}
\mathbf{W}_{\widehat{0}}=%
\begin{pmatrix}
\mathbf{E}_{\widehat{0}}
\\
\mathbf{0}%
\end{pmatrix}%
,\ \widetilde{\mathbf{W}}_{\widehat{0}}=%
\begin{pmatrix}
\mathbf{0} \\
\mathbf{E}_{\widehat{0}}
\end{pmatrix}
\ \ \mathrm{with}\ \  \mathbf{E}_{\widehat{0}}=(\delta _{\mathbf{g0}}\mathbf{I}_{3})_{\mathbf{g}\in \mathbb{Z}^{2}}.
\label{Wfur}
\end{equation}%
Calculation of $\widehat{\mathbf{S}}$ in (\ref{Cjjfur}) can be performed by
means of calculating $\mathbf{M}_{0}$ from its definition (\ref{matr2}), using
any of the known methods for evaluating multiplicative integrals. However, this
approach may suffer from numerical instabilities for $\mathbf{M}_{0}$ of
large size due to exponential growth of some components of $\mathbf{M}_{0}%
\mathbf{.}$ Another method rests on calculating the resolvent $(\mathbf{M}%
_{0}-\alpha \mathbf{I})^{-1}$ without calculating $\mathbf{M}_{0}$. The
advantage of doing so is due to the fact that growing dimension of $\mathbf{M%
}_{0}$ leads to a relatively moderate growth of the resolvent components.
Let us consider this latter method in some detail.

Denote the resolvent $\mathbf{R}_{\alpha }(x_{l})=(\mathbf{M}%
_{0}(x_{l})-\alpha \mathbf{I})^{-1}$ where $\alpha $ is not an eigenvalue of
$\mathbf{M}_{0}(x_{l})$. Since the matrix $\mathbf{M}_{0}(x_{l})$ satisfies
the differential equation $\mathbf{M}_{0}^{\prime }(x_{l})=\mathbf{Q}%
_{0}(x_{l})\mathbf{M}_{0}(x_{l})$ with the initial condition $\mathbf{M}%
_{0}(0)=\mathbf{I}$, it follows that $\mathbf{R}_{\alpha }(x_{l})$ with a
randomly chosen $\alpha \neq 1$ satisfies a Riccati equation with
initial condition as follows
\begin{equation}
\begin{cases}
\mathbf{R}_{\alpha }^{\prime }=-\mathbf{R}_{\alpha }\mathbf{Q}_{0}(\mathbf{I}%
+\alpha \mathbf{R}_{\alpha }), \\
\mathbf{R}_{\alpha }(0)=(1-\alpha )^{-1}\mathbf{I},%
\end{cases}
\label{diffres}
\end{equation}%
where $^{\prime }=\partial _{l}.$ Since eigenvalues of $\mathbf{M}_{0}$
usually tend to lie close to the real axis and unit circle, it is
recommended to take $\alpha \in {\mathbb{C}}$ far from these sets.
Integrating Eq.\ (\ref{diffres}) numerically (we used the Runge-Kutta method
of fourth order) provides $\mathbf{R}_{\alpha }(1)=(\mathbf{M}_{0}(1)-\alpha
\mathbf{I})^{-1}$ where $\alpha \neq 1.$ To exploit it for finding $\mathbf{C%
}_{ll}^{\mathrm{eff}}$ given by (\ref{Cjjfur}) we use the identity
\begin{align}
\widehat{\mathbf{S}} \equiv (\mathbf{M}_{0}(1)-\mathbf{I})^{-1}\mathbf{W}_{%
\widehat{0}}=(\mathbf{I}+(\alpha -1)\mathbf{R}_{\alpha }(1))^{-1}\mathbf{R}%
_{\alpha }(1)\mathbf{W}_{\widehat{0}}
=(\mathbf{I}+(\alpha -1)\mathbf{R}_{\alpha }(1))^{-1}\frac{\mathbf{W}_{%
\widehat{0}}}{1-\alpha }.
\end{align}%
Thus  $\widehat{\mathbf{S}}$ is found from the linear system%
\begin{equation}
(1-\alpha)
\big(\mathbf{I}+(\alpha -1)\mathbf{R}_{\alpha }(1)\big)\widehat{\mathbf{S}}=\mathbf{W}_{\widehat{0}}.  \label{R}
\end{equation}%
To solve (\ref{R}), we first note that the matrix
$\mathbf{T}\equiv (1-\alpha)
\big(\mathbf{I}+(\alpha -1)\mathbf{R}_{\alpha }(1)\big)$  satisfies $\mathbf{TW}_{\widehat{0}}=%
\mathbf{0,\ }\widetilde{\mathbf{W}}_{\widehat{0}}^{+}\mathbf{T}=\mathbf{0}$
with $\mathbf{W}_{\widehat{0}},$ $\widetilde{\mathbf{W}}_{\widehat{0}}$ from
(\ref{Wfur}) and thus has 3 zero columns on the left of its vertical midline
and 3 zero rows below the horizontal midline. Removing these columns and
rows yields the reduced matrix $\widetilde{\mathbf{T}}={\scriptsize {%
\begin{pmatrix}
\mathbf{T}_{1} & \mathbf{T}_{2} \\
\mathbf{T}_{3} & \mathbf{T}_{4}%
\end{pmatrix}%
}}$ where the square block $\mathbf{T}_{3}$ has 3 rows and 3 columns less
than the block$~\mathbf{T}_{2}.$ Since $\mathbf{T}_{3}$ is numerically
large, while $\mathbf{T}_{1},\mathbf{T}_{4}$ are medium and $\mathbf{T}_{2}$
small, it is convenient to apply Schur's formula to arrive at the final
relation in the form%
\begin{equation}
\mathbf{C}_{ll}^{\mathrm{eff}}=\mathbf{E}_{\widehat{0}%
}^{+}\big( \mathbf{T}_{2}-\mathbf{T}_{1}\mathbf{T}_{3}^{-1}\mathbf{T}%
_{4}\big)^{-1} \mathbf{E}_{\widehat{0}}
.
\label{Schur}
\end{equation}

\subsubsection{Off-principal directions}\label{3.2.3}

Consider $\mathbf{k}=k\mathbf{\kappa }$ of arbitrary orientation relatively
to the translations $\mathbf{e}_{l}.$ In order to find the effective
Christoffel tensor {$\boldsymbol{\Gamma}( \mathbf{\kappa }) $ of Eq.\ \eqref{9-0}}
for arbitrary $\mathbf{\kappa ~}$\textbf{(}$\nparallel \mathbf{e}_{l}$)$,$
we need to calculate {$\mathbf{C}_{jl}^{\mathrm{e}}$} with $j\neq l.$
Applying the 2D Fourier expansion to Eqs.\  (\ref{eq1}) and (\ref{eta1}) yields
{\begin{equation}
{\mathbf{C}_{jl}^{\mathrm{e}}}
=
\langle \mathbf{E}_{\widehat{0}}^{+} \,
( \mathbf{A}_1 +
\overline{\mathbf{C}}_{jl}\partial_{l})
\widehat{\mathbf{V}}\rangle _{l}
= \mathbf{C}_{ll}^{\mathrm{eff}} +
\langle \mathbf{E}_{\widehat{0}}^{+} \,
(
\overline{\mathbf{C}}_{ll} -\overline{\mathbf{C}}_{jl} )
\mathbf{B}^{-1} (\mathbf{A}_1
\widehat{\mathbf{V}}-
\widehat{\mathbf{N}} ) \rangle _{l},
\  \text{where} \
\begin{pmatrix}
\widehat{\mathbf{V}}
\\
\widehat{\mathbf{N}}
\end{pmatrix} = \mathbf{M}_{0}(x_{l})\widehat{\mathbf{S}}
\label{Cjlfur}
\end{equation}
and }  $\overline{\mathbf{C}}_{jp}=
(\widehat{\mathbf{C}}_{jp}
(\mathbf{g}-\mathbf{g}^{\prime},x_{l}))_{\mathbf{g},\mathbf{g}^{\prime }\in \mathbb{Z}^{2}}$ with $\mathbf{g}=\left( g_{a}\, g_{b}\right) $ and $a,b\neq l$. As detailed above, the matrix $\widehat{\mathbf{S}}$ can be
calculated with a very good precision.   Evaluation of
$\widehat{\mathbf{V}}(x_{l})$ and $\widehat{\mathbf{N}}(x_{l})$
is no longer that accurate, particularly for
high-contrast structures which require many terms in the  Fourier series and
hence need $\mathbf{M}_{0}$ of  large algebraic dimension and therefore
with large values of some components. Thus a negligible error in $\mathbf{S}$
may become noticeable after multiplying by $\mathbf{M}_{0}.$ In this regard,
we present an alternative method of calculating {$\mathbf{C}_{jl}^{\mathrm{e}}$} with $j\neq l$ which circumvents (\ref{Cjlfur}).

For the fixed functions $\rho ( \mathbf{x} ) ~$and $c_{ijkl}\left(
\mathbf{x}\right) $ with a given cubic lattice of periods $\mathbf{e}%
_{p}=\left( \delta _{pq}\right)$,  introduce the new periods $\mathbf{a}_{p}=%
\mathbf{A}{\mathbf{e}}_{p}$ where $\mathbf{A}$ has integer components
$a_{ij}$. The
solutions $\mathbf{v}(\mathbf{x})=\mathbf{u}(\mathbf{x})e^{i\mathbf{k\cdot x}%
}$ and $\omega \left( \mathbf{k}\right) $ of the wave equation (\ref{1}) remain unaltered, as they do
not  depend on the choice of periods.
{But now in
order to define
${\mathbf{C}}_{jl}^{\mathrm{eff}}$
by the MM formula (which requires
periods to coincide with base vectors) we should apply the change of variables $%
\mathbf{x\rightarrow Ax}$ that leads, as explained in \S\ref{3.4},
{to  new functions  for the density and elasticity with
the periods $\mathbf{e}_{p}=\left( \delta _{pq}\right)$.
According to Eqs.\ (\ref{transformation}) and \eqref{-00} of \S\ref{3.4},
\begin{equation}
\mathbf{C}_{pq}^{\mathrm{eff}}=
 a_{pj}\widetilde{\mathbf{C}}_{jl}^{\mathrm{eff}}a_{ql}
 = a_{pj} \mathbf{A} \overline{\mathbf{C}}_{jl}^{\mathrm{eff}}
  \mathbf{A}^+ a_{ql}
  \ \
  \Big(\Leftrightarrow \
\widetilde{\mathbf{C}}_{jl}^{\mathrm{eff}}\ =
b_{jp}\mathbf{C}_{pq}^{%
\mathrm{eff}}b_{lq} , \
\overline{\mathbf{C}}_{jl}^{\mathrm{eff}}\ =
b_{jp}\mathbf{B}\mathbf{C}_{pq}^{%
\mathrm{eff}}\mathbf{B}^+b_{lq}
\Big)
 \label{trans1}
\end{equation}
where $\widetilde{\mathbf{C}}_{jl}^{\mathrm{eff}}$
and $ \overline{\mathbf{C}}_{jl}^{\mathrm{eff}}$
are the effective
matrices associated with the new profiles $\widetilde{c}_{ijkl}( \mathbf{x} ) = b_{jp}b_{lq} c_{ipkq}( \mathbf{A}\mathbf{x}) $
and $\overline{c}_{ijkl}( \mathbf{x} ) = b_{im}b_{jp}b_{kn}b_{lq} c_{mpnq}( \mathbf{A}\mathbf{x}) $, respectively,
and
$b_{ij}$ stand for components of $\mathbf{A}^{-1}$.
}
 We may apply
(\ref{trans1}) successively for each of the  transformations
$\mathbf{a}_{p}=\mathbf{A}_{j}{\mathbf{e}}_{p}$, $j=1,2,3$,
{
\small
\begin{equation}
\mathbf{A}_{1} =
\begin{pmatrix}
1 &0&0 \\
0& 1&-1 \\
0& 1&1
\end{pmatrix}
 \Leftrightarrow
 \begin{aligned}
\mathbf{a}_{1}&= \mathbf{e}_{1},\\
\mathbf{a}_{2}&= \mathbf{e}_{2}+\mathbf{e}_{3},\\
\mathbf{a}_{3}&= -\mathbf{e}_{2}+\mathbf{e}_{3},
\end{aligned}
\ \
\mathbf{A}_{2} =
\begin{pmatrix}
1 &0&1 \\
0& 1&0 \\
-1& 0&1
\end{pmatrix}
 \Leftrightarrow  ...,
\
\mathbf{A}_{3} =
\begin{pmatrix}
1 &-1&0 \\
1& 1&0 \\
0& 0&1
\end{pmatrix}
  \Leftrightarrow  ... \
.
\label{Ak}
\end{equation}}
{Using the inverse forms in (\ref{trans1})  leads to a variety of identities, for instance,
{\begin{equation}
\begin{aligned}
\mathbf{C}_{23}^{\mathrm{eff}}+\mathbf{C}_{32}^{\mathrm{eff}}&=4(\widetilde{\mathbf{C}}_{22}^{\mathrm{eff}})_{\mathbf{A}_{1}} - \mathbf{C}_{22}^{\mathrm{eff}}-\mathbf{C}_{33}^{\mathrm{eff}}
=4(\mathbf{A}\overline{\mathbf{C}}_{22}^{\mathrm{eff}}\mathbf{A}^+)_{\mathbf{A}_{1}} -\mathbf{C}_{22}^{\mathrm{eff}}-\mathbf{C}
_{33}^{\mathrm{eff}},
\\
\mathbf{C}_{31}^{\mathrm{eff}}+\mathbf{C}_{13}^{\mathrm{eff}}&=4(\widetilde{\mathbf{C}}%
_{33}^{\mathrm{eff}})_{\mathbf{A}_{2}}-\mathbf{C}_{33}^{\mathrm{eff}}-\mathbf{C}_{11}^{\mathrm{eff}}
=4(\mathbf{A}\overline{\mathbf{C}}%
_{33}^{\mathrm{eff}}\mathbf{A}^+)_{\mathbf{A}_{2}}
-\mathbf{C}_{33}^{\mathrm{eff}}-\mathbf{C}_{11}^{\mathrm{eff}},
 \\
\mathbf{C}_{12}^{\mathrm{eff}}+\mathbf{C}_{21}^{\mathrm{eff}}&=4(\widetilde{\mathbf{C}}%
_{11}^{\mathrm{eff}})_{\mathbf{A}_{3}} -\mathbf{C}_{11}^{\mathrm{eff}}-\mathbf{C}_{22}^{\mathrm{eff}}
=4(\mathbf{A}\overline{\mathbf{C}}%
_{11}^{\mathrm{eff}}\mathbf{A}^+)_{\mathbf{A}_{3}}-\mathbf{C}_{11}^{\mathrm{eff}}+\mathbf{C}_{22}^{\mathrm{eff}}.
\end{aligned}
\label{off}
\end{equation}}
Inserting (\ref{off}) in {(\ref{9-0})} eliminates $\mathbf{C}_{jl}^{\mathrm{eff}%
}+\mathbf{C}_{lj}^{\mathrm{eff}}$ with $j\neq l$ and expresses the effective
Christoffel tensor $\boldsymbol{\Gamma}$ fully in terms of the matrices $\mathbf{C}%
_{ll}^{\mathrm{eff}}$ and either of  $(\widetilde{\mathbf{C}}_{ll}^{\mathrm{eff}})_{%
\mathbf{A}_{n}}$
or $(\overline{\mathbf{C}}_{ll}^{\mathrm{eff}})_{%
\mathbf{A}_{n}}$  ($n=1,2,3$), which are defined by Eqs.\  (\ref{hh3}), (\ref%
{Cjjfur}) with $ c_{ijkl}\left( \mathbf{x}%
\right) $ replaced  by    $\widetilde{c}_{ijkl}\left(\mathbf{x}\right)$
or  $\overline{c}_{ijkl}\left(\mathbf{x}\right)$, respectively. Thus all calculations have been reduced to the form (\ref{Cjjfur}) which ensures a  numerically stable  evaluation of $\boldsymbol{\Gamma}$.
}

{
Note that the transformed elasticity
 $\overline{c}_{ijkl}( \mathbf{x} ) $ retains the usual symmetries under the interchange of indices, while
  $\widetilde{c}_{ijkl}( \mathbf{x} ) $ does not (see \S\ref{3.4}).
  Also,  the transformations defined by \eqref{Ak} for the cubic unit cell can be identified as rotations by virtue of the fact that
  $\frac 1{\sqrt{2}}\mathbf{A}_{j}$, $ j=1,2,3$, are orthogonal matrices of unit determinant.    Hence, apart from a factor of $\frac 14$,
  $\big(\overline{c}_{ijkl}( \mathbf{x} )\big)_{\mathbf{A}_{j}}$ is precisely the elasticity tensor represented in a coordinate system rotated about the axis $\mathbf{e}_j$ by $\frac{\pi}4$ from the original.
}}


\subsection{The case of an oblique lattice}\label{3.4}

\subsubsection{Equivalent problem on a cubic lattice}

Consider the problem of quasistatic asymptotics of the wave equation
\eqref{1}
for the general case of a 3D periodic elastic medium with
\begin{equation}
\rho ( \mathbf{x}) =\rho \left( \mathbf{x+a}_{p}\right) ,\
c_{ijkl}( \mathbf{x}) =c_{ijkl}\left( \mathbf{x+a}_{p}\right) ,
\label{-1'}
\end{equation}%
where the translation vectors $\mathbf{a}_{p}$ form an oblique lattice. We
will define the solution of this problem via the solution for a simpler case
of a cubic lattice.

{The oblique lattice vectors are defined by a matrix $\mathbf{A}$
$(\ne \mathbf{I})$ as }
\begin{equation}
\mathbf{a}_{p}=\mathbf{Ae}_{p}\ \mathrm{with}\ \mathbf{A}=\left(
a_{pq}\right) _{p,q=1}^{3}=\left( \mathbf{a}_{p}\cdot \mathbf{e}_{q}\right)
_{p,q=1}^{3};\ \mathbf{B} \equiv \mathbf{A}^{-1}=\left( b_{pq}\right) _{p,q=1}^{3}  \label{A}
\end{equation}%
where $\mathbf{e}_{p}$ are the orthonormal vectors used previously.
Define the new or transformed position variable
$\mathbf{x}' = \mathbf{B} \mathbf{x}$  $(\Leftrightarrow
\mathbf{x} = \mathbf{A} \mathbf{x}^\prime )$, the associated
displacement $\widetilde{\mathbf{v}}  (\mathbf{x}')
= {\mathbf{v}}  (\mathbf{x})$ and material parameters
  $ \widetilde{\rho } (\mathbf{x}') = \rho (\mathbf{x}) $,
{$  c_{ijkl}^{(1)}(\mathbf{x}') = c_{ijkl}(\mathbf{x}) $,}
which are seen to be periodic in $\mathbf{x}'$ with respect to the  vectors $\mathbf{e}_{p}$.   Setting
{${\mathbf{C}}_{jl}^{(1)}
=\big( {c}_{ijkl}^{(1)}\big)_{i,k=1}^{3}$}, the equation of motion \eqref{1} becomes
\begin{equation}
{
b_{jp} b_{lq}\,
\partial _{j'}\big(
{ {\mathbf{C}}_{pq}^{(1)} }\partial _{l'} \widetilde{\mathbf{v}} \big)}=-\widetilde{\rho }\omega ^{2}\widetilde{\mathbf{v}},  \label{09}
\end{equation}%
where $\partial _{j'} \equiv \partial /\partial x_{j}'$.  Using the fact that  $\mathbf{B}$ is constant  allows it to be removed explicitly from \eqref{09} by incorporation into a newly defined stiffness tensor.  Thus, replacing $ \mathbf{x}' \to \mathbf{x}$ we have
\begin{equation}
\partial _{j}(\widetilde{\mathbf{C}}_{jl}\partial _{l}\widetilde{\mathbf{v}}%
)=-\widetilde{\rho }\omega ^{2}\widetilde{\mathbf{v}},  \label{3dweq1}
\end{equation}%
where $\widetilde{\mathbf{v}}(\mathbf{x})=\mathbf{v}(\mathbf{Ax})$
and the  material parameters are
\begin{equation}
\begin{aligned}
\widetilde{\rho }(\mathbf{x})
&=\rho (\mathbf{Ax})\ \ \left( =\widetilde{\rho }%
(\mathbf{x+e}_{p})\right) , \\
\widetilde{c}_{ijkl}(\mathbf{x})
&=b_{jp}b_{lq}c_{ipkq}(\mathbf{Ax})\ \ \left(
=\widetilde{c}_{ijkl}(\mathbf{x+e}_{p})\right) , \\
\widetilde{\mathbf{C}}_{jl}(\mathbf{x})
&=\left( \widetilde{c}_{ijkl}\right)
_{i,k=1}^{3}=b_{jp}b_{lq}\mathbf{C}_{pq}(\mathbf{Ax})=\widetilde{\mathbf{C}}%
_{lj}^{+}(\mathbf{x}),%
\end{aligned}
\label{replace}
\end{equation}%
which are periodic in $\mathbf{x}$ with respect to the cubic lattice formed by
vectors $\mathbf{e}_{p}$. Note that the tensor $\widetilde{c}_{ijkl}$ for  $%
\mathbf{A\neq I}$ is of Cosserat type in that it is not invariant to
permutations of indices $i\rightleftarrows j$ and $k\rightleftarrows l$ but
retains the major symmetry $\widetilde{c}_{ijkl}=\widetilde{c}_{klij}^{\ast }~(=%
\widetilde{c}_{klij}$ for real case)$.$

 The
Floquet condition $\mathbf{v}(\mathbf{x})=e^{i\mathbf{k}\cdot \mathbf{x}}%
\mathbf{u}(\mathbf{x})$ with periodic $\mathbf{u}(\mathbf{x})=\mathbf{u}(%
\mathbf{x+a}_{j})$   satisfying 
\begin{equation}
-(\partial _{l}+ik_{l})\mathbf{C}_{lq}(\partial _{q}+ik_{q})\mathbf{u}=\rho
\omega ^{2}\mathbf{u}  \label{eqo1}
\end{equation}%
is equivalent to the condition $%
\widetilde{\mathbf{v}}(\mathbf{x})
=e^{i  \widetilde{\mathbf{k}}\cdot \mathbf{x} }
\widetilde{\mathbf{u}}(\mathbf{x})
$
with periodic $\widetilde{%
\mathbf{u}}(\mathbf{x})$ satisfying the equation that follows from \eqref{09},
\begin{equation}
-(\partial _{l}+i\widetilde{k}_{l})\widetilde{\mathbf{C}}_{lq}(\partial
_{q}+i\widetilde{k}_{q})\widetilde{\mathbf{u}}=\widetilde{\rho }\widetilde{%
\omega }^{2}\widetilde{\mathbf{u}},  \label{eqo2}
\end{equation}%
where
\begin{equation}
\widetilde{\mathbf{k}}=\mathbf{A}^+ \mathbf{k\ (}=\widetilde{k}%
\widetilde{\mathbf{\kappa }},\ \left\vert \widetilde{\mathbf{\kappa }}%
\right\vert =1),\ \ \widetilde{\omega }(\widetilde{\mathbf{k}})=\omega (%
\mathbf{k}),\ \ \widetilde{\mathbf{u}}(\mathbf{x})=\mathbf{u}(\mathbf{A}%
\mathbf{x})\ \left( =\widetilde{\mathbf{u}}(\mathbf{x+e}_{p})\right) .
\label{parameters}
\end{equation}
Equation (\ref{eqo2}), which is defined on a cubic lattice, has quasistatic
asymptotics as described above. According to (\ref{eq}),
\begin{equation}
\widetilde{\boldsymbol{\Gamma}}\widetilde{\mathbf{\mathbf{u}}}_{0\alpha }=\langle
\widetilde{\rho }\rangle \widetilde{c}_{\alpha }^{2}\widetilde{\mathbf{%
\mathbf{u}}}_{0\alpha }\ \mathrm{with}\ \widetilde{c}_{\alpha }(\widetilde{%
\boldsymbol{\kappa }})\equiv \lim_{\widetilde{k}\rightarrow 0}\widetilde{\omega }_{\alpha }(%
\widetilde{\mathbf{k}})/\widetilde{k},\ \alpha =1,2,3,  \label{eff1}
\end{equation}%
where {$\widetilde{\mathbf{\Gamma }}(\widetilde{\boldsymbol{\kappa }}%
\mathbf{)}=\widetilde{\kappa }_{j}\widetilde{\kappa }_{l}\widetilde{\mathbf{C%
}}_{jl}^{\mathrm{eff}}$}.
Let us write a similar relation for the  quasistatic
asymptotics of (\ref{eqo1}),
\begin{equation}
{\boldsymbol{\Gamma}}\mathbf{\mathbf{u}}_{0\alpha }=\langle {\rho }\rangle
c_{\alpha }^{2}\mathbf{\mathbf{u}}_{0\alpha }\ \mathrm{with}\ c_{\alpha }(%
\boldsymbol{\kappa })\equiv \lim_{k\rightarrow 0}\omega _{\alpha }(\mathbf{k})/k,
\label{eff2}
\end{equation}%
where $\mathbf{k}=k\mathbf{\boldsymbol{\kappa }}$ $\mathbf{(}\left\vert
\boldsymbol{\kappa
}\right\vert =1)$ and ${\boldsymbol{\Gamma}}(\boldsymbol{\kappa })$ is to be determined.
Comparing (\ref{eff1}) and (\ref{eff2}) with regard for (\ref{parameters})
and making use of the equality
{$\langle \widetilde{\rho }\rangle \equiv
\int_{\left[ 0,1\right] ^{3}}\rho ( \mathbf{A}\mathbf{x}) d\mathbf{x}%
= 
\langle \rho \rangle $}, we find
that
\begin{equation}
\frac{1}{\langle \rho \rangle }k^{2}{\boldsymbol{\Gamma}=}\frac{1}{\langle
\widetilde{\rho }\rangle }\widetilde{k}^{2}\widetilde{\boldsymbol{\Gamma}}\
\Rightarrow \ {\boldsymbol{\Gamma}}=\frac{\langle \rho \rangle }{\langle \widetilde{%
\rho }\rangle }\frac{\widetilde{k}_{j}\widetilde{k}_{l}}{k^{2}}\widetilde{%
\mathbf{C}}_{jl}^{\mathrm{eff}}=
{\kappa_{p}a_{pj}\widetilde{\mathbf{C}}_{jl}^{\mathrm{eff}}a_{ql}{\kappa_{q}}
}
\end{equation}%
and hence
\begin{equation}
{\boldsymbol{\Gamma}}(\boldsymbol{\kappa })=\mathbf{C}_{pq}^{\mathrm{eff}}\kappa
_{p}\kappa _{q}:\ \ \mathbf{C}_{pq}^{\mathrm{eff}}=
{ a_{pj}\widetilde{\mathbf{C}}_{jl}^{\mathrm{eff}}a_{ql}
\ \ \Big(\Leftrightarrow \
\widetilde{\mathbf{C}}_{jl}^{\mathrm{eff}}\ =
b_{jp}\mathbf{C}_{pq}^{%
\mathrm{eff}}b_{lq} \Big).
}
\label{transformation}
\end{equation}

\subsubsection{Alternative formulation using anisotropic mass density}

{
Premultiplication of Eq.\ \eqref{3dweq1} by $\mathbf{B}$ allows it to be reformulated as
\begin{equation}\label{-8}
\partial _{j}(\overline{\mathbf{C}}_{jl}\partial _{l}\overline{\mathbf{v}}%
)=-\overline{\boldsymbol \rho }\omega ^{2}\overline{\mathbf{v}},
\end{equation}%
where $\overline{\mathbf{v}}$
and the  material parameters are
\begin{equation}\label{2-0}
\begin{aligned}
\overline{\mathbf{v}} (\mathbf{x})&= \mathbf{A}^+  \mathbf{v}(\mathbf{Ax})
\ \ \big( = \mathbf{A}^+ \widetilde{\mathbf{v}}(\mathbf{x}) \big),
\\
\overline{\boldsymbol \rho }(\mathbf{x})
&=\mathbf{B}\mathbf{B}^+ \rho (\mathbf{Ax})
=\mathbf{B}\mathbf{B}^+ \widetilde{\rho} (\mathbf{x})
= \overline{\boldsymbol \rho }^+(\mathbf{x}) \ \ \left( =\overline{\rho }%
(\mathbf{x+e}_{p})\right) , \\
\overline{c}_{ijkl}(\mathbf{x})
&=b_{im}b_{jp}b_{kn}b_{lq}c_{mpnq}(\mathbf{Ax})\ \ \left(
=\overline{c}_{ijkl}(\mathbf{x+e}_{p})\right) , \\
\overline{\mathbf{C}}_{jl}(\mathbf{x})
&=\left( \overline{c}_{ijkl}\right)
_{i,k=1}^{3}=b_{jp}b_{lq}\mathbf{B}\mathbf{C}_{pq}(\mathbf{Ax})\mathbf{B}^+
=\mathbf{B}\widetilde{\mathbf{C}}_{pq}(\mathbf{x})\mathbf{B}^+ =\overline{\mathbf{C}}%
_{lj}^{+}(\mathbf{x}),%
\end{aligned}
\end{equation}%
which are periodic in $\mathbf{x}$ with respect to the cubic lattice of
vectors $\mathbf{e}_{p}$. Note that the tensor $\overline{c}_{ijkl}$
retains the major and minor symmetries of normal elasticity,  $\overline{c}_{ijkl}=\overline{c}_{klij}^{\ast }$ and
$\overline{c}_{ijkl}=\overline{c}_{jikl}$,
while the mass density is no longer a scalar but becomes a symmetric tensor.
}

{
 The
Floquet condition now becomes  $\overline{\mathbf{v}}(\mathbf{x})
=e^{i  \overline{\mathbf{k}}\cdot \mathbf{x} }
\overline{\mathbf{u}}(\mathbf{x})$
with periodic $\overline{
\mathbf{u}}(\mathbf{x})$ satisfying
\begin{equation}
-(\partial _{l}+i\overline{k}_{l})\overline{\mathbf{C}}_{lq}(\partial
_{q}+i\overline{k}_{q})\overline{\mathbf{u}}=\overline{\boldsymbol\rho }\, \overline{\omega }^{2}\overline{\mathbf{u}},  \label{-6}
\end{equation}%
where
\begin{equation}
\overline{\mathbf{k}}=\mathbf{A}^+ \mathbf{k\ (}=\overline{k}%
\overline{\mathbf{\kappa }},\ \left\vert \overline{\mathbf{\kappa }}%
\right\vert =1),\ \ \overline{\omega }(\overline{\mathbf{k}})=\omega (%
\mathbf{k}),\ \ \overline{\mathbf{u}}(\mathbf{x})=\mathbf{A}^+\mathbf{u}(\mathbf{A}%
\mathbf{x})\ \left( =\overline{\mathbf{u}}(\mathbf{x+e}_{p})\right) .
\label{-77}
\end{equation}
Its  quasistatic
asymptotics are
\begin{equation}
\overline{\boldsymbol{\Gamma}}\overline{\mathbf{\mathbf{u}}}_{0\alpha }=\langle
\overline{\boldsymbol\rho }\rangle \overline{c}_{\alpha }^{2}\overline{\mathbf{%
\mathbf{u}}}_{0\alpha }\ \mathrm{with}\ \overline{c}_{\alpha }(\overline{%
\boldsymbol{\kappa }})\equiv \lim_{k\rightarrow 0}\overline{\omega }_{\alpha }(%
\overline{\mathbf{k}})/\overline{k},\ \alpha =1,2,3,  \label{-98}
\end{equation}%
where $\overline{\mathbf{\Gamma }}(\overline{\boldsymbol{\kappa }}%
\mathbf{)}=\overline{\kappa }_{j}\overline{\kappa }_{l}\overline{\mathbf{C%
}}_{jl}^{\mathrm{eff}}$.
Comparing (\ref{eff2}) and (\ref{-98}) with regard for (\ref{-77})
and making use of the equality
$\langle \overline{\boldsymbol\rho } \rangle
 = \mathbf{B}\mathbf{B}^+ \langle\widetilde{\rho }\rangle
= \mathbf{B}\mathbf{B}^+
\langle \rho \rangle $, we find
that
\begin{equation}
\langle \rho \rangle ^{-1} k^{2}{\boldsymbol{\Gamma}=}\langle
{\rho }\rangle^{-1} \overline{k}^{2}
\mathbf{A}\overline{\boldsymbol{\Gamma}}\mathbf{A}^+\
\Rightarrow \ {\boldsymbol{\Gamma}}
=
{\kappa_{p}a_{pj}\mathbf{A}\overline{\mathbf{C}}_{jl}^{\mathrm{eff}}\mathbf{A}^+a_{ql}{\kappa }_{q}
}
\end{equation}%
and hence
\begin{equation}
 \mathbf{C}_{pq}^{\mathrm{eff}}=
{ a_{pj}\mathbf{A}\overline{\mathbf{C}}_{jl}^{\mathrm{eff}}
\mathbf{A}^+a_{ql}
\ \ \Big(\Leftrightarrow \
\overline{\mathbf{C}}_{jl}^{\mathrm{eff}}\ =
b_{jp}\mathbf{B}\mathbf{C}_{pq}^{%
\mathrm{eff}}\mathbf{B}^+b_{lq} \Big).
}
\label{-00}
\end{equation}
}

{
\subsubsection{Summary of the oblique case}
Given material constants $\rho ( \mathbf{x}) $ and $%
c_{ijkl}( \mathbf{x}) $ on an oblique lattice we first
identify the matrix $\mathbf{A}$  of (\ref{A}). We
may proceed in either of two ways based on
 Cosserat elasticity with isotropic density, or normal elasticity with anisotropic density.  In each case
we define material properties  on a cubic lattice:
$\widetilde{\rho }( \mathbf{x})$,
 $\widetilde{\mathbf{C}}_{jl}( \mathbf{x}) $ from Eq.\ \eqref{replace} or $\overline{\boldsymbol \rho }( \mathbf{x})$,  $\overline{\mathbf{C}}_{jl}( \mathbf{x}) $ from Eq.\ \eqref{2-0}, respectively. Then
use the formulas of \S3 to obtain $\widetilde{\mathbf{C}}_{jl}( \mathbf{x}) \to \widetilde{\mathbf{C}}_{jl}^{\mathrm{eff}}$ or  $\overline{\mathbf{C}}_{jl}( \mathbf{x})  \to \overline{\mathbf{C}}_{jl}^{\mathrm{eff}} $, and finally
insert the result into (\ref{transformation}) or \eqref{-00} to arrive at the sought
effective Christoffel matrix ${\boldsymbol{\Gamma}}$ as a function of unit
direction vector $\boldsymbol{\kappa }$ in the oblique lattice.
 Knowing ${%
\boldsymbol{\Gamma}(}\boldsymbol{\kappa })$ yields the effective speeds $c_{\alpha }(%
\boldsymbol{\kappa })$ according to (\ref{eff2}).
Note that although the formulation in \S2  was restricted to isotropic density, the  quasi-static effective elasticity
is  the same if one replaces the isotropic density tensor $\rho (\mathbf{x}) \mathbf{I}$ by the anisotropic density
$\rho (\mathbf{x}) \mathbf{J}$ with $\mathbf{J} = \mathbf{J}^+$ constant positive definite. In the case of the anisotropic density formulation for the oblique lattice $\mathbf{J} = \mathbf{B}\mathbf{B}^+$.
}

{
\subsection{Calculating the effective elastic moduli.} \label{full}

{We return to the question of the full determination of
$c_{ijkl}^{\mathrm{eff}}$ from the  Christoffel tensor, or more
specifically, from  $\mathbf{D}$ defined by
  \begin{equation} \label{009}
d_{ikjl} =
\frac 12
 \big(
c_{ijkl}^{\mathrm{eff}}
+c_{ilkj}^{\mathrm{eff}}\big)
.
\end{equation}
The  elements of $\mathbf{D}$ satisfy the same symmetries as those of $\mathbf{C}$
$(d_{ikjl} = d_{jlik} = d_{iklj})$ and they follow from Eq.\ \eqref{123} as
\begin{equation} \label{321}
(d_{ikjl})_{i,k=1}^{3} =
\frac 12 \big( \mathbf{C}_{jl}^{\mathrm{eff}} +\mathbf{C}_{lj}^{\mathrm{eff}}\big)  =
\frac 12 \big(\mathbf{C}_{jl}^{\mathrm{e}} +\mathbf{C}_{lj}^{\mathrm{e}} \big)
\equiv \mathbf{D}_{jl} \ \ \big( =\mathbf{D}_{lj}\big)
.
\end{equation}
}
Define the  'totally symmetric' part of $\mathbf{C}^{\mathrm{eff}}$ as
$c_{ijkl}^{\mathrm{eff,s}}
= \frac 13\big( c_{ijkl}^{\mathrm{eff}}
+c_{ikjl}^{\mathrm{eff}} +c_{iljk}^{\mathrm{eff}} \big)
$.   
This is seen to be equal to the
totally symmetric part of $\mathbf{D}$ defined in  \eqref{009}, i.e.
$\mathbf{C}^{\mathrm{eff,s}}=\mathbf{D}^{\mathrm{s}}$ where
$d_{ijkl}^{\mathrm{s}} = \frac 13\big( d_{ijkl}
+d_{ikjl}  +d_{iljk}  \big)$.  Equation \eqref{009}
can then be rewritten  as \cite{Norris05g}
\begin{equation}\label{2=7}
\mathbf{D} = \frac 32 \mathbf{C}^{\mathrm{eff,s}}
- \frac 12 \mathbf{C}^{\mathrm{eff}}
\ \Rightarrow \
\mathbf{C}^{\mathrm{eff}} = 3\mathbf{D}^{\mathrm{s}} - 2 \mathbf{D}.
\end{equation}
  The latter inverse relation may be simply represented  in Voigt notation as
\begin{align}\label{1=4}
\begin{pmatrix}
c_{11}^{\mathrm{eff}} &
c_{12}^{\mathrm{eff}} &
c_{13}^{\mathrm{eff}} &
c_{14}^{\mathrm{eff}} &
c_{15}^{\mathrm{eff}} &
c_{16}^{\mathrm{eff}} \\
  &
c_{22}^{\mathrm{eff}} &
c_{23}^{\mathrm{eff}} &
c_{24}^{\mathrm{eff}} &
c_{25}^{\mathrm{eff}} &
c_{26}^{\mathrm{eff}} \\
  & &
c_{33}^{\mathrm{eff}} &
c_{34}^{\mathrm{eff}} &
c_{35}^{\mathrm{eff}} &
c_{36}^{\mathrm{eff}} \\
  & & &
c_{44}^{\mathrm{eff}} &
c_{45}^{\mathrm{eff}} &
c_{46}^{\mathrm{eff}} \\
  S& Y& M& &
c_{55}^{\mathrm{eff}} &
c_{56}^{\mathrm{eff}} \\
  & & & & &
c_{66}^{\mathrm{eff}}
\end{pmatrix}
=
\begin{pmatrix}
d_{11} &
2d_{66}-d_{12} &
2d_{55}-d_{13} &
2d_{56}-d_{14} &
d_{15} &
d_{16} \\
  &
d_{22} &
2d_{44}-d_{23} &
d_{24} &
2d_{46}-d_{25} &
d_{26} \\
  & &
d_{33} &
d_{34} &
d_{35} &
2d_{45}-d_{36} \\
  & & &
d_{23} &
d_{36} &
d_{25} \\
  & SYM& & &
d_{13} &
d_{14} \\
  & & & & &
d_{12}
\end{pmatrix}.
\end{align}
This one-to-one correspondence between the elements
$\mathbf{C}^{\mathrm{eff}}$ and $\mathbf{D} $, combined with the identity \eqref{321}, provides the means to find the effective moduli from $\mathbf{C}_{jl}^{\mathrm{e}}$.
}

{
It remains to determine the full set of  elements $d_{ijkl}$ from Christoffel tensors for a given set of directions.
  It is known that data
for at least six distinct directions  are required \cite{cowin86,norris89}.  The necessary and sufficient condition that a given  sextet
$\{{\boldsymbol{\Gamma}}^{(\alpha)} \equiv {\boldsymbol{\Gamma}}(
{\boldsymbol \kappa}^{\alpha}) , \alpha =1,.., 6\}$ will yield the full elastic moduli tensor is that the six  directions
$\{{\boldsymbol \kappa}^{\alpha} \}$  do not lie on a cone through the origin and cannot be contained in less than three distinct planes through the origin \cite{norris89}.
The set   $\{{\boldsymbol \kappa}^{\alpha} \} = \{ \mathbf{e}_1, \mathbf{e}_2, \mathbf{e}_3,
\frac 1{\sqrt{2}}\mathbf{A}_{1}{\mathbf{e}_2},
\frac 1{\sqrt{2}}\mathbf{A}_{2}{\mathbf{e}_3},
\frac 1{\sqrt{2}}\mathbf{A}_{3}{\mathbf{e}_1}
\}$ (see \eqref{Ak}) meets this requirement (and is in fact the set first proposed in \cite{cowin86}).
Thus the complete set of  $d_{ijkl}$ follows from \cite[Eq.\ (3.17)]{norris89}  (see Eq.\ \eqref{321})
\begin{align}\label{2=5}
\mathbf{D}_{jl}
=&\sum\limits_{\alpha=1}^3
\boldsymbol{\Gamma}^{(\alpha)}
\big( \kappa^{\alpha}_j\kappa^{\alpha}_l
-\tfrac 1{\sqrt{2}} ( \kappa^{\alpha +3}_j\kappa^{\alpha}_l
+\kappa^{\alpha}_j\kappa^{\alpha +3}_l ) \big)
+ \sum\limits_{\stackrel{\alpha , \beta =1}{\beta > \alpha}}^3
\boldsymbol{\Gamma}^{(9-\alpha -\beta)}
( \kappa^{\alpha}_j\kappa^{\beta}_l
+ \kappa^{\beta}_j\kappa^{\alpha}_l ) .
\end{align}
The  equivalent form of \eqref{2=5} in Voigt notation is
\begin{align}\label{1=3}
\begin{pmatrix}
d_{11} &
d_{12} &
d_{13} &
d_{14} &
d_{15} &
d_{16} \\
  &
d_{22} &
d_{23} &
d_{24} &
d_{25} &
d_{26} \\
  & &
d_{33} &
d_{34} &
d_{35} &
d_{36} \\
  & & &
d_{44} &
d_{45} &
d_{46} \\
  S& Y&M & &
d_{55} &
d_{56} \\
  & & & & &
d_{66}
\end{pmatrix}
=
\begin{pmatrix}
\Gamma_{11}^{(1)} &
\Gamma_{11}^{(2)} &
\Gamma_{11}^{(3)} &
\Gamma_{11}^{(4)} &
\Gamma_{11}^{(5)} &
\Gamma_{11}^{(6)} \\
\Gamma_{22}^{(1)} &
\Gamma_{22}^{(2)} &
\Gamma_{22}^{(3)} &
\Gamma_{22}^{(4)} &
\Gamma_{22}^{(5)} &
\Gamma_{22}^{(6)} \\
\Gamma_{33}^{(1)} &
\Gamma_{33}^{(2)} &
\Gamma_{33}^{(3)} &
\Gamma_{33}^{(4)} &
\Gamma_{33}^{(5)} &
\Gamma_{33}^{(6)} \\
\Gamma_{23}^{(1)} &
\Gamma_{23}^{(2)} &
\Gamma_{23}^{(3)} &
\Gamma_{23}^{(4)} &
\Gamma_{23}^{(5)} &
\Gamma_{23}^{(6)} \\
\Gamma_{31}^{(1)} &
\Gamma_{31}^{(2)} &
\Gamma_{31}^{(3)} &
\Gamma_{31}^{(4)} &
\Gamma_{31}^{(5)} &
\Gamma_{31}^{(6)} \\
\Gamma_{12}^{(1)} &
\Gamma_{12}^{(2)} &
\Gamma_{12}^{(3)} &
\Gamma_{12}^{(4)} &
\Gamma_{12}^{(5)} &
\Gamma_{12}^{(6)}
\end{pmatrix}
\begin{pmatrix}
1 & 0& 0& 0&  -\frac 12 &  -\frac 12 \\
0 & 1& 0&  -\frac 12 &0&   -\frac 12 \\
0& 0 & 1&   -\frac 12 &   -\frac 12 &0 \\
0& 0 &0 & 1&0 &0 \\
0& 0 &0 & 0&1 &0 \\
0& 0 &0 & 0&0 &1
\end{pmatrix}.
\end{align}
The full set of $c_{ijkl}^{\mathrm{eff}}$ can  therefore be obtained  from Eqs.\
\eqref{1=4}  and \eqref{1=3}  with (see Eq.\ \eqref{off})
\begin{equation}\label{4=4}
\begin{aligned}
\boldsymbol{\Gamma}^{(1)} &= {\mathbf{C}}_{11}^{\mathrm{eff}},
\
&\boldsymbol{\Gamma}^{(2)}& = {\mathbf{C}}_{22}^{\mathrm{eff}},
\
&\boldsymbol{\Gamma}^{(3)}& = {\mathbf{C}}_{33}^{\mathrm{eff}},
\\
\boldsymbol{\Gamma}^{(4)} &= 2(\mathbf{A}\overline{\mathbf{C}}_{22}^{\mathrm{eff}}\mathbf{A}^+)_{\mathbf{A}_{1}},
\
 &\boldsymbol{\Gamma}^{(5)} &= 2(\mathbf{A}\overline{\mathbf{C}}_{33}^{\mathrm{eff}}\mathbf{A}^+)_{\mathbf{A}_{2}},
\
&\boldsymbol{\Gamma}^{(6)} &= 2(\mathbf{A}\overline{\mathbf{C}}_{11}^{\mathrm{eff}}\mathbf{A}^+)_{\mathbf{A}_{3}}.
\end{aligned}
\end{equation}
Other procedures for inverting a set of  compatible Christoffel tensors to give the moduli can be found in \cite{norris89}.
}

\section{Closed-form upper bound of the effective Christoffel tensor}\label{sec4}

Let $N\geq 0$ be the truncation parameter of the 3D or 2D Fourier expansion
of $\mathbf{C}_{jl}( \mathbf{x}) $, meaning that the index $%
\mathbf{g}$ in (\ref{fur3}) or (\ref{fur2}) takes the values from the set $%
[-N,N]^{3}$ or $[-N,N]^{2},$ respectively. Denote truncated approximations
of the effective Christoffel tensor $\mathbf{\Gamma }\left( \mathbf{\kappa }%
\right) =\mathbf{C}_{jl}^{\mathrm{eff}}\kappa _{j}\kappa _{l}$ calculated
from \eqref{em}$_1$ via the PWE and MM formulas by $\mathbf{\Gamma }\left[ N%
\right] _{\mathrm{PWE}}$ and $\mathbf{\Gamma }\left[ N\right] _{\mathrm{MM}}$%
, respectively. By analogy with \cite{KSN1}, it can be proved that%
\begin{equation}
\begin{aligned}
\ N_{1}\leq N_{2}\ &\Rightarrow \mathbf{\Gamma }\left[ N_{1}\right] _{\mathrm{%
MM}}\geq \mathbf{\Gamma }\left[ N_{2}\right] _{\mathrm{MM}},\ \mathbf{\Gamma
}\left[ N_{1}\right] _{\mathrm{PWE}}\geq \mathbf{\Gamma }\left[ N_{2}\right]
_{\mathrm{PWE}}; \\
\forall N\ &\Rightarrow \mathbf{\Gamma }\leq \mathbf{\Gamma }\left[ N\right]
_{\mathrm{MM}}\leq \mathbf{\Gamma }\left[ N\right] _{\mathrm{PWE}},%
\end{aligned}
\label{b0}
\end{equation}%
where the inequality sign between two matrices is understood in the sense
that their difference is a sign definite matrix and hence the differences of
their similarly ordered eigenvalues are sign definite. The inequalities  (\ref{b0})$_{1}$ imply that
the MM and PWE approximations of $\mathbf{\Gamma }$ obtained by truncating
Eq.\ \eqref{em}$_1$ are upper bounds which converge from above to the exact value
with growing $N$. Furthermore,    the MM approximation of $\mathbf{%
\Gamma }$ is more accurate than the PWE one at a given $N$, from (\ref{b0})$_{2}$.

Taking $N=0$ in (\ref{b0})$_{2}$
yields
\begin{equation}
\mathbf{\Gamma }\leq \mathbf{\Gamma }\left[ 0\right] _{\mathrm{MM}}\leq
\mathbf{\Gamma }\left[ 0\right] _{\mathrm{PWE}}=\left\langle \mathbf{\Gamma }%
\right\rangle ,  \label{b1}
\end{equation}%
where $\mathbf{\Gamma }\left[ 0\right] _{\mathrm{MM}}$ admits an explicit
expression which is however rather cumbersome. Seeking a simpler result,
consider the above inequalities for one of the principal directions $\mathbf{%
\kappa \parallel e}_{l}$ so that $\mathbf{\Gamma }\left( \mathbf{\kappa }%
\right) =\mathbf{C}_{ll}^{\mathrm{eff}}.$ Denote the PWE and MM
approximations (\ref{PWEF}) and (\ref{Cjjfur}) of $\mathbf{C}_{ll}^{\mathrm{%
eff}}$ by $\mathbf{C}_{ll}^{\mathrm{eff}}\left[ N\right] _{\mathrm{PWE}}$
and $\mathbf{C}_{ll}^{\mathrm{eff}}\left[ N\right] _{\mathrm{MM}}.$ From (%
\ref{b1}),
\begin{equation}
\mathbf{C}_{ll}^{\mathrm{eff}}\leq \mathbf{C}_{ll}^{\mathrm{eff}}\left[ 0%
\right] _{\mathrm{MM}}\leq \mathbf{C}_{ll}^{\mathrm{eff}}\left[ 0\right] _{%
\mathrm{PWE}}=\left\langle \mathbf{C}_{ll}\right\rangle ,  \label{b2}
\end{equation}%
where $\mathbf{C}_{ll}^{\mathrm{eff}}\left[ 0\right] _{\mathrm{MM}}$ admits
closed-form expression as follows. From (\ref{matrQ}) and (\ref{matr2}) taken
with $N=0$ (i.e. with $\mathbf{g,g}^{\prime }=\mathbf{0}$),
\begin{equation}
\mathbf{Q}_{0} \left[ 0\right]
=\begin{pmatrix}
\mathbf{0} & \langle \mathbf{C}_{ll}\rangle _{\overline{l}}^{-1} \\
\mathbf{0} & \mathbf{0}%
\end{pmatrix}%
\ \Rightarrow \
\mathbf{M}_{0}(1)
=
\begin{pmatrix}
\mathbf{I}_{3} & \langle \langle \mathbf{C}_{ll}\rangle _{\overline{l}%
}^{-1}\rangle _{l} \\
\mathbf{0} & \mathbf{I}_{3}%
\end{pmatrix}%
,  \label{b3}
\end{equation}%
so that (\ref{Cjjfur}) with  $N=0$ yields
\begin{equation}
\mathbf{C}_{ll}^{\mathrm{eff}}[ 0] _{\mathrm{MM}}
=\big(
\mathbf{0} \ \  \mathbf{I}_{3}
\big) \mathbf{S}[ 0] \equiv \mathbf{S}_{2}\ \ \ \mathrm{with}
\ \
\mathbf{S}[ 0]
=%
\begin{pmatrix}
\mathbf{0} & \langle \langle \mathbf{C}_{ll}\rangle _{\overline{l}%
}^{-1}\rangle _{l} \\
\mathbf{0} & \mathbf{0}%
\end{pmatrix}%
^{-1}%
\begin{pmatrix}
\mathbf{I}_{3} \\
\mathbf{0}%
\end{pmatrix}%
\equiv
\begin{pmatrix}
\mathbf{S}_{1} \\
\mathbf{S}_{2}%
\end{pmatrix}%
.%
   \label{b4}
\end{equation}%
Solving for $\mathbf{S}_{2}~$gives
\begin{equation}
\begin{pmatrix}
\mathbf{0} & \langle \langle \mathbf{C}_{ll}\rangle _{\overline{l}%
}^{-1}\rangle _{l} \\
\mathbf{0} & \mathbf{0}%
\end{pmatrix}%
\begin{pmatrix}
\mathbf{S}_{1} \\
\mathbf{S}_{2}%
\end{pmatrix}%
=%
\begin{pmatrix}
\mathbf{I}_{3} \\
\mathbf{0}%
\end{pmatrix}%
\ \Rightarrow \mathbf{S}_{2}=\langle \langle \mathbf{C}_{ll}\rangle _{%
\overline{l}}^{-1}\rangle _{l}^{-1}.  \label{b5}
\end{equation}%
Thus from (\ref{b2}), (\ref{b4}) and (\ref{b5}),
\begin{equation}
\mathbf{C}_{ll}^{\mathrm{eff}}\leq \langle \langle \mathbf{C}_{ll}\rangle _{%
\overline{l}}^{-1}\rangle _{l}^{-1}\leq \langle \mathbf{C}_{ll}\rangle
\label{b6}
\end{equation}
{where $\langle \mathbf{C}_{ll}\rangle$ is identifiable as the Voigt
average \cite{Milton01},  known to provide an upper bound.  The
upper bound provided by the first inequality in \eqref{b6} has not
to our knowledge been presented before.
}

{Combining the new bound from \eqref{b6} with the Voigt inequality $
\mathbf{C}_{al}^{\mathrm{eff}}\leq \langle \mathbf{C}_{al}\rangle $
$(a\ne l)$ yields }
\begin{equation}
\mathbf{\Gamma }( \mathbf{\kappa }) \leq \mathbf{\Gamma }_{%
\mathrm{B}}( \mathbf{\kappa }) \equiv
\left( \langle \mathbf{C}_{al}\rangle \kappa _{a}\kappa _{l}\right) _{a\neq l}+\langle \langle
\mathbf{C}_{ll}\rangle _{\overline{l}}^{-1}\rangle _{l}^{-1}\kappa _{l}^{2},
\label{b7}
\end{equation}
where the subscript "B" implies bound. Denote the eigenvalues of the matrix $%
\mathbf{\Gamma }_{\mathrm{B}}$ by $\left\langle \rho \right\rangle c_{%
\mathrm{B}\alpha }^{2}$ and order them in the same way as the eigenvalues $%
\left\langle \rho \right\rangle c_{\alpha }^{2}$ of $\mathbf{\Gamma }$, then
it follows that
\begin{equation}
c_{\alpha }( \mathbf{\kappa }) \leq c_{\mathrm{B}\alpha }(
\mathbf{\kappa }) ,\ \alpha =1,2,3.  \label{b8}
\end{equation}%
It will be demonstrated in \S\ref{sec5} that the upper bounds $c_{\mathrm{B}\alpha
}$ of the effective speeds can also serve as their reasonable estimate.

\begin{figure}[H]
\begin{minipage}[h]{0.4\linewidth}
\center{\includegraphics[width=1\linewidth]{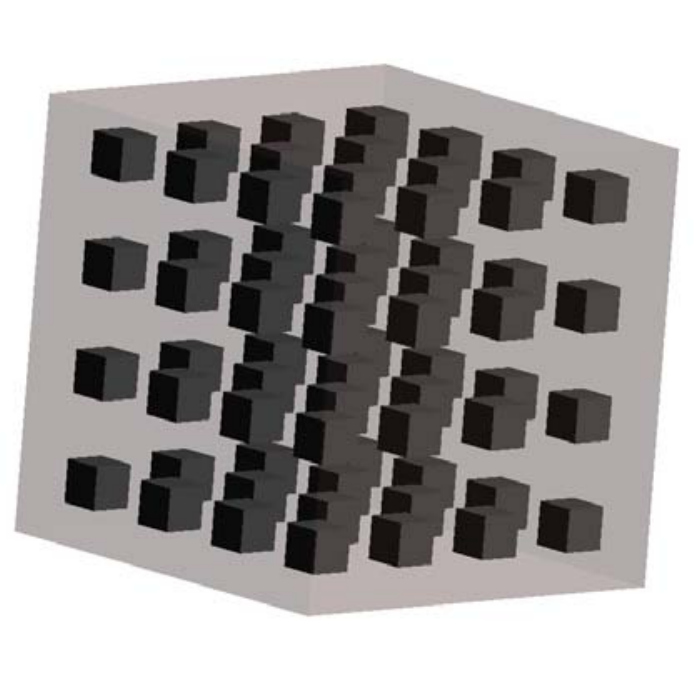} \\ a)}
\end{minipage}
\hfill
\begin{minipage}[h]{0.59\linewidth}
\center{\includegraphics[width=1\linewidth]{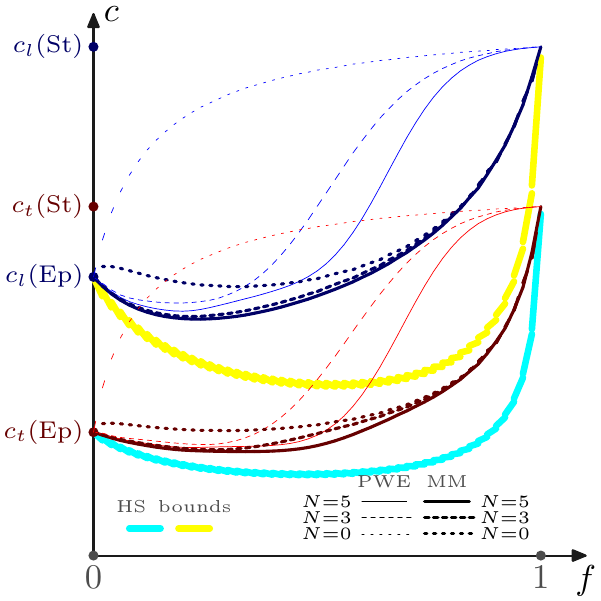} \\ b)}
\end{minipage}
\caption{(a) A cubic lattice of symmetric steel cubes in
Epoxy at filling fraction $f= 1/8$.  (b) Effective wave speeds as a function of $f$.
The PWE and MM calculated values are  plotted
by thin and thick lines (light blue and red online), respectively.   The broad curves indicate the Hashin-Shtrikman lower bounds.
} \label{fig1}
\end{figure}

\section{Numerical examples }\label{sec5}
We consider two examples of 3D phononic crystals composed of steel
inclusions in epoxy matrix. The material parameters are $c_{11}\left(
\mathrm{St}\right) =170$ GPa, $c_{66}\left( \mathrm{St}\right) =80$ GPa,~$%
\rho \left( \mathrm{St}\right) =7.7$ g/cm$^{3}$ for steel and $c_{11}\left(
\mathrm{Ep}\right) =7.537$ GPa, $c_{66}\left( \mathrm{Ep}\right) =1.482$
GPa,~$\rho \left( \mathrm{Ep}\right) =1.142$ g/cm$^{3}$ for epoxy. This
implies $c_{l}\left( \mathrm{St}\right) =4.7$ mm/%
$\mu$%
s, $c_{t}\left( \mathrm{St}\right) =3.22$ mm/%
$\mu$%
s and $c_{l}\left( \mathrm{Ep}\right) =2.57$ mm/%
$\mu$%
s, $c_{t}\left( \mathrm{Ep}\right) =1.14$ mm/%
$\mu$%
s for the longitudinal and transverse speeds. The number of Fourier modes is
$\left( 2N+1\right) ^{3}$ for the PWE method and $\left( 2N+1\right) ^{2}$
for the MM method; we performed the calculations for $N=0,3,5$.

The first
example assumes a cubic lattice of cubic steel inclusions (Fig.\ \ref{fig1}a). We
present the effective longitudinal and transverse speeds $c_{l}$ and $c_{t}$
in the principal direction as functions of the volume fraction of steel
inclusions (Fig.\ \ref{fig1}b). The curves calculated by the PWE method are plotted
by thin lines (light blue and red online), the curves calculated by the MM
method are plotted by thick lines (dark blue and red online). For each
method, we present three different data obtained with $N=0,~N=3$ and $N=5$
(dotted, dashed and solid lines, respectively). The Hashin-Shtrikman lower bounds \cite{Hashin63} are also plotted (the Hashin-Shtrikman upper bounds lie  far above the other curves and are not displayed).  It is observed from Fig.\ \ref{fig1}b
that the results of both methods monotonically converge from above to the
exact value with growing $N$ in agreement with the general statement of \S\ref{sec4}. What is significant is that the convergence of the MM method is seen to
be much faster than that of the PWE method. In fact, the explicit MM
estimate for $N=0$ which follows from \eqref{b6} in the form
\begin{equation}\label{bound}
c_{l}^{2}=\frac{1}{\left\langle \rho \right\rangle }\left\langle
\left\langle c_{11}\right\rangle _{\overline{1}}^{-1}\right\rangle
_{1}^{-1},\ c_{t}^{2}=\frac{1}{\left\langle \rho \right\rangle }\left\langle
\left\langle c_{66}\right\rangle _{\overline{1}}^{-1}\right\rangle _{1}^{-1},
\end{equation}
provides a much better estimate for $c_{l}$ and $c_{t}$ at $f>0.5$
than the PWE calculation with $N=5,$ i.e. with matrices of about
4000$\times $4000 size. { Note that as $f\to 1$ in the example of
Fig.\ \ref{fig1},
the bound
\eqref{b6} may be approximated, yielding
\begin{equation}
c_{11}^{\mathrm{eff}}
\lessapprox
\Big( \frac 1{ c_{11}(\mathrm{St})}
+ \frac {1-f^{1/3}}{ c_{11}(\mathrm{Ep})}
\Big)^{-1} .
\label{b61}
\end{equation}
At the same time the geometry of the unit cell for  $f\to 1$
indicates that the  modulus  $c_{11}^{\mathrm{eff}}$ can be estimated
by an equivalent medium stratified in the $1-$direction, for which
the uniaxial strain  assumption with constant stress $\sigma_{11}$ leads to the approximation
$c_{11}^{\mathrm{eff}} \approx \langle c_{11}^{-1} \rangle^{-1}$,   the same as the right hand side of \eqref{b61} (as $f\to 1)$.   This, combined with the fact that $1-f^{1/3}$ tends to zero faster than $1-f$ as $f\to 1$, explains the exceptional accuracy of the new bound as an estimate for the moduli.  Note that, by comparison,  the Hashin-Shtrikman bounds (upper or lower) do not provide a useful estimate in this case.
}

\begin{figure}[H]
\begin{minipage}[h]{0.49\linewidth}
\center{\includegraphics[width=1\linewidth]{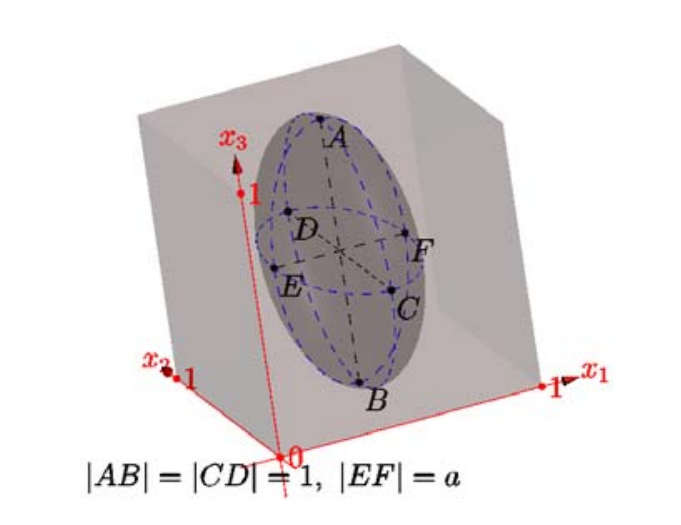} \\ a)}
\end{minipage}
\hfill
\begin{minipage}[h]{0.49\linewidth}
\center{\includegraphics[width=1\linewidth]{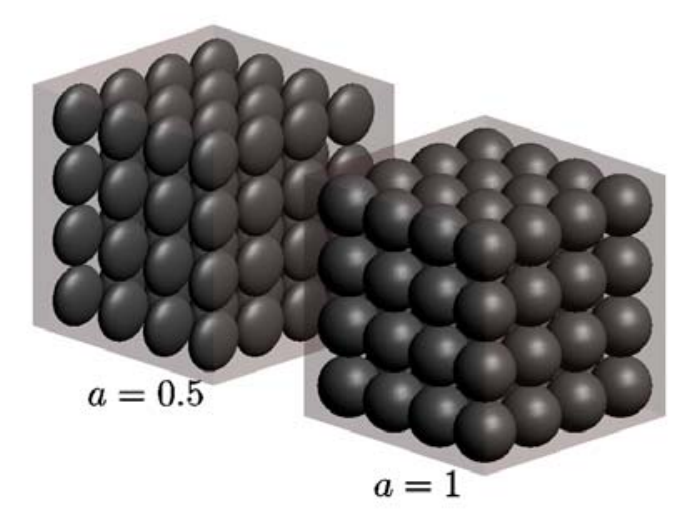} \\ b)}
\end{minipage}
\caption{A cubic lattice of  Steel  spheroids in
Epoxy matrix. (a) The inclusions are oblate spheroids with minor axis $a$ and unit major axes.  (b) The periodic structure for $a=0.5$ and $a=1$ (spheres).  } \label{fig2}
\end{figure}

\begin{figure}[H]
\begin{minipage}[h]{0.49\linewidth}
\center{\includegraphics[width=1\linewidth]{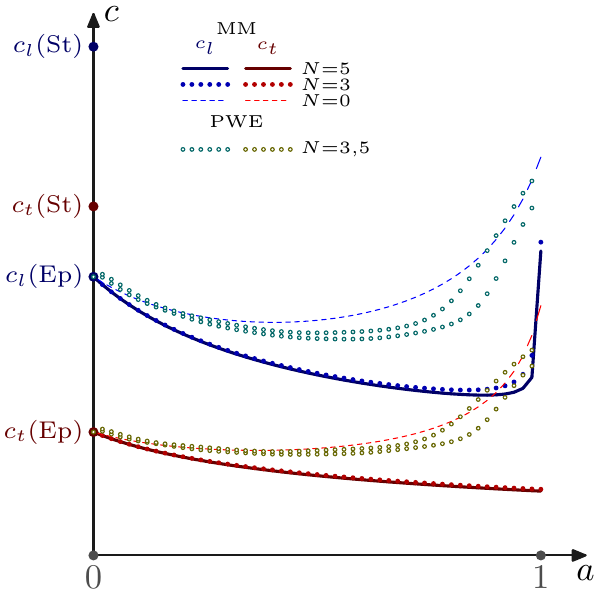} \\ a)}
\end{minipage}
\hfill
\begin{minipage}[h]{0.49\linewidth}
\center{\includegraphics[width=1\linewidth]{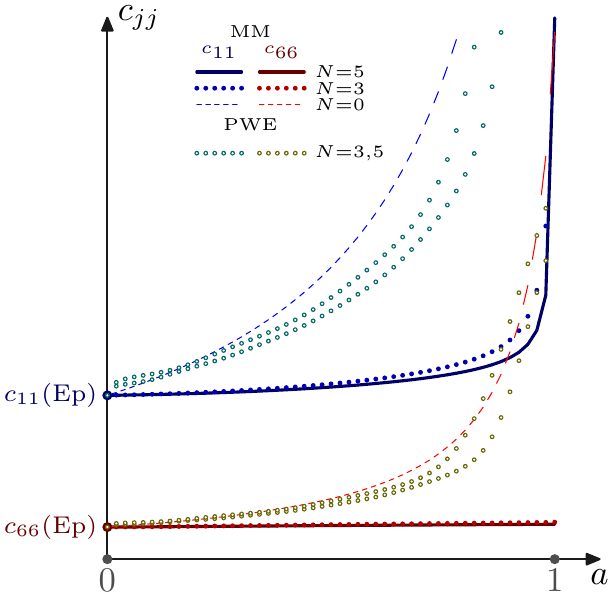} \\ b)}
\end{minipage}
\caption{(a) Effective wave speeds for the periodic structure of Fig.\ \ref{fig2} as a function of the spheroid minor axis $a$ calculated by the PWE and MM methods.  (b) The corresponding elastic moduli. } \label{fig3}
\end{figure}

The second example considers a cubic lattice of spheroidal steel inclusions
in epoxy matrix. The shape of the  inclusion evolves from formally a
disk of unit diameter to a ball of unit diameter (that is, inscribed in
a cubic unit cell) by means of elongating the radius along the $x_{1}$
direction, see Fig.\ \ref{fig2}. We describe the dependence of the effective speeds $%
c_{l}$ and $c_{t}$ along $x_{1}$ and of the corresponding effective
elastic moduli $c_{11}$ and $c_{66}$ on the shape of the spheroidal
inclusion. Fourier coefficients for the PWE and MM methods are given
in Appendix 2.  The results are obtained by the PWE method with
$N=3$ and $N=5$ (open circles in Fig.\ \ref{fig3}) and by the MM
method with $N=0,~N=3$ and $N=5$ (dotted, dashed and solid lines in
Fig.\ \ref{fig3}). The MM method is particularly efficient for the
case in hand since it uses Fourier coefficients in the $x_{2}x_{3}$
plane where the inclusions have circular cross-section and performs
direct numerical integration of the Riccati equation (see
\S\ref{sec3.2}) along the direction $x_{1}$ where the shape is
'distorted'. We observe an interesting feature of a drastic increase
of the effective longitudinal speed $c_{l}$ and of the modulus
$c_{11}$ when the inclusions tend to touch each other, see Fig.\
\ref{fig3}a.    This type of   configuration where the inclusions
are almost touching is known to be particularly difficult  for
numerical calculation of the effective properties \cite{Nunan84}. MM
appears to be particularly well suited to treating such problems
with closely spaced inclusions since it explicitly accounts for the
thin gap region via integration of the Riccati equation. { The PWE
method, on the other hand, clearly fails to capture the sharp
increase in wave speed  at $N=5$ (matrix size $\approx$ 4000$\times
$4000).  In fact, the
 PWE for $N=3$ does not even satisfy  the strict upper bounds \eqref{bound} derived from MM at $N=0$.
 }

\section{Conclusion}\label{sec6}

The PWE and MM methods of calculating quasistatic effective speeds
in three-dimensional phononic crystals have been formulated and compared. The MM method can be viewed as a two-dimensional PWE combined with a one-dimensional propagator matrix approach.  The propagator part of the MM scheme  is calculated by numerical integration of a (nonlinear) Riccati differential equation to produce the monodromy matrix.

It was shown both analytically and numerically that  the MM method provides more accurate approximations than the PWE scheme. In particular, the closed form MM bounds \eqref{b6} (see also \eqref{bound}) using only one Fourier mode
to estimate the  effective speed gives better approximations than PWE
bounds using more than a thousand (eleven in each of $x_i$, $i=1,2,3$) Fourier
modes in the case of densely packed structures (see Fig.\ \ref{fig1}b for $f > 0.5$).

 The speed-up of the MM method as compared with PWE via reduction in matrix size is particularly significant for the three-dimensional homogenization problem.  Thus,
numerical implementation of the PWE scheme needs a  matrix of dimension $3(2N+1)^3\times3(2N+1)^3$, requiring a considerable amount  of computer memory even for small $N$.  By contrast, the
MM scheme uses  matrices of
dimension $6(2N+1)^2\times6(2N+1)^2$.  The reduced  memory requirement for the MM method is at the cost of the computer time needed to solve the Riccati equation, a relatively small price to pay.  In fact, the ability to set the step size in the Runge-Kutta scheme enables the MM method to efficiently and accurately solve configurations for which the PWE is particularly ill-suited, such as narrow gaps (see Fig.\ \ref{fig1}b for $f\to1$) and closely spaced inclusions (Fig.\ \ref{fig3} for $a\to1$).


\section*{Appendix}

\subsection*{Appendix 1. Alternative derivation of Eq.\ (\ref{hh3}) for $\mathbf{C}_{ll}^{\mathrm{eff}}$.}

Let $\mathbf{k\ }$be parallel to one of the translation vectors. Take the
latter to be $\mathbf{e}_{1}=\left( \delta _{1i}\right) $ and so $\mathbf{k}%
=(k_{1},0,0).$ Equation (\ref{1}) may be rewritten in the form
\begin{align} \label{qq0}
\boldsymbol{\eta }^{\prime }
=(\boldsymbol{\mathcal{Q}}_{0}+\omega ^{2}\boldsymbol{\mathcal{Q}}_{1})\boldsymbol{\eta }%
\ \ \mathrm{\ where\ }^{\prime }\equiv \partial _{1},  \ \ 
\boldsymbol{\eta}
=%
\begin{pmatrix}
\mathbf{v} \\
\mathbf{C}_{1p}\partial _{p}\mathbf{v}%
\end{pmatrix}%
,\ \boldsymbol{\mathcal{Q}}_{1}=%
\begin{pmatrix}
\mathbf{0} & \mathbf{0} \\
-\rho \mathbf{I}_{3} & \mathbf{0}%
\end{pmatrix}%
,  
\end{align}
while $\boldsymbol{\mathcal{Q}}_{0}$ is defined in (\ref{hh0}) and (\ref{hh2}) with $j=1\
$and $a,b=2,3.$ Denote $\boldsymbol{\eta}\left( x_{1}\right) \equiv \boldsymbol{\eta}%
(x_{1},x_{2},x_{3})$. The solution to (\ref{qq0}) with some initial function
$\boldsymbol{\eta}\left( 0\right) $ can be written via the matricant in the form
\begin{equation}
\boldsymbol{\eta}\left( x_{1}\right) =\boldsymbol{\mathcal{M}}(x_{1})\boldsymbol{\eta}\left( 0\right) \
\ \mathrm{with\ }\boldsymbol{\mathcal{M}}(x_{1})=\widehat{\int_{0}^{x_{1}}}(\boldsymbol{\mathcal{I}}+(%
\boldsymbol{\mathcal{Q}}_{0}+\omega ^{2}\boldsymbol{\mathcal{Q}}_{1})dx_{1}).  \label{mon}
\end{equation}%
Taking into account assumed 1-periodicity in $x_{1}$ and hence the Floquet
condition $\mathbf{v}=e^{ik_{1}x_{1}}\mathbf{u}$ for the solution of (\ref{1}%
) implies that the solution $\boldsymbol{\eta}$ of (\ref{qq0}) must satisfy $%
\boldsymbol{\eta}(1)=e^{ik_{1}}\boldsymbol{\eta}(0)$. Thus, with reference to (\ref{mon}),
$\omega (k_{1},0,0)\equiv \omega (k_{1})$ is an eigenvalue of (\ref{1}) iff
there exists $\mathbf{w}\equiv \mathbf{w}(x_{2},x_{3})$ such that
\begin{equation}
\boldsymbol{\mathcal{M}}(1)\mathbf{w}=e^{ik_{1}}\mathbf{w}.  \label{M}
\end{equation}%
Consider asymptotic expansion of (\ref{M}) in small $\omega ,k_{1}.$ By (\ref%
{mon}),
\begin{equation} \label{M6}
\begin{aligned}
\boldsymbol{\mathcal{M}}(1)&=\boldsymbol{\mathcal{M}}_{0}+\omega ^{2}\boldsymbol{\mathcal{M}}_{1}+O(\omega ^{4})\ \
\mathrm{where}    \\
\boldsymbol{\mathcal{M}}_{0}&\equiv \boldsymbol{\mathcal{M}}_{0}\left[ 1,0\right] ,\ \ \boldsymbol{\mathcal{M}}_{0}%
\left[ b,a\right] =%
\begin{matrix}
\widehat{\int }_{a}^{b}%
\end{matrix}%
(\boldsymbol{\mathcal{I}}+\boldsymbol{\mathcal{Q}}_{0}\mathrm{d}x_{1}),  \ \
\boldsymbol{\mathcal{M}}_{1}  
=\int_{0}^{1}\boldsymbol{\mathcal{M}}_{0}\left[ 1,x_{1}\right] \boldsymbol{\mathcal{Q}}%
_{1}\boldsymbol{\mathcal{M}}_{0}\left[ x_{1},0\right] \mathrm{d}x_{1}.
\end{aligned}
\end{equation}%
The identity $\boldsymbol{\mathcal{M}}_{0}\mathbf{W}_{0}=\mathbf{W}_{0}$ with the $%
6\times 3$ matrix $\mathbf{W}_{0}=\left( \mathbf{I}_{3}\ \mathbf{0}\right)^+$ (see (\ref{i})) implies triple multiplicity of the zero-order $%
\omega =0$. Therefore we may write
\begin{equation} \label{asw2}
\begin{aligned}
\omega _{\alpha }(k_{1})=c_{\alpha }k_{1}+O(k_{1}^{2}),\ \ \
\mathbf{w}_{\alpha }=\mathbf{w}_{0\alpha }+k_{1}\mathbf{w}_{1\alpha
}+k_{1}^{2}\mathbf{w}_{2\alpha }+\mathbf{O}(k_{1}^{3})\ \mathrm{with}\
\mathbf{w}_{0\alpha }=\mathbf{W}_{0}\mathbf{u}_{0\alpha },  
\end{aligned}
\end{equation}%
where $\alpha =1,2,3$ and $\mathbf{u}_{0\alpha }$ are some constant linear
independent $3\times 1$ vectors. Inserting (\ref{M6})-(\ref{asw2}) along
with $e^{ik_{1}}=1+ik_{1}-\frac{1}{2}k_{1}^{2}+O(k_{1}^{3})$ in (\ref{M})
and equating the terms of the same order in $k_{1}$ yields%
\begin{subequations}
\begin{align}
1 :&\ \boldsymbol{\mathcal{M}}_{0}\mathbf{w}_{0\alpha }=\mathbf{w}_{0\alpha }, \\
k_{1} :& \ \boldsymbol{\mathcal{M}}_{0}\mathbf{w}_{1\alpha }=i\mathbf{w}_{0\alpha }+\mathbf{%
w}_{1\alpha },  \label{other1} \\
k_{1}^{2} :& \ \boldsymbol{\mathcal{M}}_{0}\mathbf{w}_{2\alpha }+c_{\alpha }^{2}\boldsymbol{\mathcal{M}}%
_{1}\mathbf{w}_{0\alpha }=-\frac{1}{2}\mathbf{w}_{0\alpha }+i\mathbf{w}%
_{1\alpha }+\mathbf{w}_{2\alpha }.  \label{other2}
\end{align}%
\end{subequations}
Express $\mathbf{w}_{1\alpha }$ from (\ref{other1}) and substitute it in (%
\ref{other2}), then scalar multiply the latter by the $6\times 3$ matrix $%
\widetilde{\mathbf{W}}_{0}=\left( \mathbf{0}\
\mathbf{I}_{3}\right)^{+} $ satisfying the identity
$\boldsymbol{\mathcal{M}}^+_{0}\left[ b,a\right]
\widetilde{\mathbf{W}}_{0}=\widetilde{\mathbf{W}}_{0}$ (see
(\ref{i})). As a result, we obtain
\begin{equation}
\mathbf{C}_{11}^{\mathrm{eff}}=\langle \widetilde{\mathbf{W}}_{0}^{+}(%
\boldsymbol{\mathcal{M}}_{0}-\boldsymbol{\mathcal{I}})^{-1}\mathbf{W}_{0}\rangle _{\overline{1}}\ \
\mathrm{for}\ \mathbf{\kappa =e}_{1}=\left( \delta _{1i}\right) .
\label{res11}
\end{equation}%
It is seen that (\ref{hh3}) with $\boldsymbol{\mathcal{M}}_{0}\left( 1\right) \equiv
\boldsymbol{\mathcal{M}}_{0}$ and $l=1$ is the same as (\ref{res11}), QED.

\subsection*{Appendix 2.   Fourier coefficients for spheroidal inclusions}

The coefficients for the spheroids of Fig. 2a are as follows:

{\bf 1. MM method.} Identity (27) yields
\[\label{001}
 \widehat{\bf C}_{pq}(g_2,g_3,x_1)=\begin{cases} {\bf C}_{pq}({\rm Ep}), & x_1\not\in\left[\frac{1-a}2,\frac{1+a}2\right],\\
                                   {\bf C}_{pq}({\rm Ep})+({\bf C}_{pq}({\rm St})-{\bf C}_{pq}({\rm
                                   Ep}))\hat\chi_1(g_2,g_3,x_1), & x_1\in\left[\frac{1-a}2,\frac{1+a}2\right]
                                   \end{cases}
\]
with
\[\label{002}
 \hat\chi_1(g_2,g_3,x_1)=(-1)^{g_2+g_3}\frac{R J_1(2\pi R\sqrt{g_2^2+g_3^2})}{\sqrt{g_2^2+g_3^2}},\ \
 R^2=1-\frac{(2x_1-1)^2}{a^2},
\]
where $J_1$ is  the first order Bessel function.

{\bf 2. PWE method.} Identity (13) yields
\[\label{003}
 \widehat{\bf C}_{pq}({\bf g})={\bf C}_{pq}({\rm Ep})\delta_{{\bf g} {\bf 0}}+({\bf C}_{pq}({\rm St})-{\bf C}_{pq}({\rm Ep}))\hat\chi_2({\bf
 g}),
\]
where
\[\label{004}
 \hat\chi_2({\bf
 g})=\frac{a(-1)^{g_1+g_2+g_3}}{2\pi^2|{\bf g}_a|^3}(\sin(\pi|{\bf g}_a|)-\pi|{\bf g}_a|\cos(\pi|{\bf
 g}_a|)),\ \ |{\bf g}_a|=\sqrt{(ag_1)^2+g_2^2+g_3^2}
\]
and $\delta$ is a Kronecker symbol.

\subsection*{Acknowledgment}

A.A.K. acknowledges support from Mairie de Bordeaux.
A.N.N. acknowledges support from Institut de M\'{e}canique et d'Ing\'{e}nierie, Universit\'{e} de Bordeaux.

%

\end{document}